\documentclass[useAMS,usenatbib]{mn2e}
\usepackage{amssymb}
\usepackage{graphicx}
\usepackage{lscape}
\usepackage{rotating}
\usepackage{color}
\usepackage{float}

\newcommand{\chandra}{{\it Chandra}\,}

\newcommand{\xmm}{{\em XMM--Newton}\,}

\newcommand{\bc}{\begin{center}}
\newcommand{\ec}{\end{center}}
\def\ltsima{$\; \buildrel < \over \sim \;$}
\def\lsim{\lower.5ex\hbox{\ltsima}}
\def\loe{\lower.5ex\hbox{\ltsima}}
\def\gtsima{$\; \buildrel > \over \sim \;$}
\def\gsim{\lower.5ex\hbox{\gtsima}}
\def\goe{\lower.5ex\hbox{\gtsima}}

\def\ltsima{$\; \buildrel < \over \sim \;$}
\def\lsim{\lower.5ex\hbox{\ltsima}}
\def\loe{\lower.5ex\hbox{\ltsima}}
\def\gtsima{$\; \buildrel > \over \sim \;$}
\def\gsim{\lower.5ex\hbox{\gtsima}}
\def\goe{\lower.5ex\hbox{\gtsima}}

\def\ergscm2 {erg\,s$^{-1}$cm$^{-2}$}

\def\cm2 {cm$^{-2}$}

\def\ergscm2 {erg\,s$^{-1}$cm$^{-2}$}

\def\ee {1E\,1048.1--5937\,}

\def\cxo {CXOU\,010043-7211\,}
\def\sgra{SGR\,1900+14\,}
\def\sgrb{SGR\,1806--20\,}

\begin{document}

\title[Comparing SNRs around PSRs]
{Comparing supernova remnants around strongly magnetized and canonical pulsars}
\author[Martin, Rea, Torres \& Papitto]{J. Martin$^{1}$, N. Rea$^{1,2}$, D. F. Torres$^{1,3}$ \& A. Papitto$^{1}$ \\
$^1$ Institute of Space Sciences (CSIC--IEEC), Campus UAB, Faculty of Science, Torre C5-parell, E-08193 Barcelona, Spain \\
$^2$ Anton Pannekoek Institute for Astronomy, University of Amsterdam, Postbus 94249,  NL-1090-GE Amsterdam, The Netherlands\\
$^3$ Instituci\'o Catalana de Recerca i Estudis Avanats (ICREA), 08010 Barcelona, Spain}
\date{}
\maketitle

\label{firstpage}

\begin{abstract}
  
The origin of the strong magnetic fields measured in magnetars is one of the main uncertainties in the neutron star field. On the other hand, the
recent discovery of a large number of such strongly magnetized neutron stars, is calling for more investigation on their formation. The first
proposed model for the formation of such strong magnetic fields in magnetars was through alpha-dynamo effects on the rapidly rotating core of a
massive star. Other scenarios involve highly magnetic massive progenitors that conserve their strong magnetic moment into the core after the
explosion, or a common envelope phase of a massive binary system. In this work, we do a complete re-analysis of the archival X-ray emission of the Supernova Remnants (SNR) surrounding magnetars, and
compare our results with all other bright X-ray emitting SNRs, which are associated with Compact Central Objects (CCOs; which are
proposed to have magnetar-like B-fields buried in the crust by strong accretion soon after their formation), high-B pulsars and normal pulsars. We find that emission lines in SNRs
hosting highly magnetic neutron stars do not differ significantly in elements or ionization state from those observed in other SNRs, neither
averaging on the whole remnants, nor studying different parts of their total spatial extent. Furthermore, we find no significant evidence that the total X-ray luminosities of SNRs hosting magnetars, are on average larger than that of typical young X-ray SNRs. Although biased by a small number of objects, we found that for a similar age, there is the same percentage of magnetars showing a detectable SNR than for the normal pulsar population.

\end{abstract}

\section{Introduction}\label{intro}

Supernova explosions are among the most energetic and extreme events ever observed in the Universe. Supernovae are mainly distinguished in two
main classes: core-collapse (CC) and thermonuclear supernovae. Core-collapse SNe result from the core collapse of a massive star
($>8$ M$_{\odot}$; see \citealt{woosley05}, for a review), while thermonuclear SNe are due to the explosion of a white dwarf in a binary system
with a giant star (single-degenerate origin), or from two low-mass white dwarfs in a binary system (double-degenerate origin; \citealt{hille00}).
Core-collapse SNe might leave behind a fast rotating (several milliseconds) and strongly magnetized ($>10^{12}$ G) stellar core which is now made by
degenerate matter: a so-called neutron star. At the same time the envelope of the massive star, ejected at high speed ($\sim10^4$km\,s$^{-1}$) into
the interstellar medium, interacts with it, resulting in what is called a Supernova Remnant (SNR). In the standard picture a SNR evolves in time
following four main expansion phases: free expansion, Sedov-Taylor phase, radiative and merging phase. The timescales and properties of each of
those phases are characterized by the initial SN explosion energy, interstellar ambient density, and the age of the remnant (see \citealt{vink12}
for a recent review).

In the recent years, a class of highly magnetized neutron stars (a.k.a. magnetars) have been discovered. Magnetars 
 are a small group of X-ray pulsars (about twenty objects with spin periods between 2--12 s) the emission of
which is not explained by the common scenario for pulsars. In fact, the very strong X-ray emission of these objects
(L$_{x}\sim10^{35}$ erg) seemed too high and variable to be fed by the rotational energy alone (as in the radio pulsars), and no evidence for a
companion star has been found in favor of any accretion process (see \citealt{mere08} and \citealt{rea11} for reviews). Assuming the typical
magnetic loss equation for rotating neutron stars, their inferred magnetic fields appear to be in general of the order of
$B \sim 10^{14} - 10^{15}$ G (although low magnetic field magnetars have been recently discovered \citealt{rea10,rea12}).
Because of these high B fields, the emission of magnetars is thought to be powered by the decay and the instability of their strong fields
\citep{duncan92,thompson93,thompson02}.

The exact mechanism playing a key role in the formation of such strong magnetic fields is currently debated; in particular it is not clear which
are the characteristics of a massive star turning into a``magnetar'' instead of a normal radio pulsar, after its supernova explosion.

Preliminary calculations have shown that the effects of a turbulent dynamo amplification occurring in a newly born neutron star can indeed
result in a magnetic field of a few $10^{17}$ G. This dynamo effect is expected to operate only in the first $\sim10$ s after the supernova
explosion of the massive progenitor, and if the proto-neutron star is born with sufficiently small rotational periods (of the order of a few ms).
The resulting amplified magnetic fields are expected to have a strong multipolar structure, and toroidal component
\citep{duncan92,duncan96,thompson93}. However, this scenario is encountering more and more difficulties: i) if magnetic torques can indeed remove
angular momentum from the core via the coupling to the atmosphere in a pre-SN phase, then the core soon after the SN might not spin rapidly
enough for this convective dynamo mechanism to take place \citep{heger05}; ii) such a fast spinning proto-neutron star would require a supernova
explosion one order of magnitude more energetic than normal supernovae, possibly an hypernova, which is not yet clear on whether it can indeed
form a neutron star instead of a black hole. Recent simulations have shown that gamma-ray bursts (GRBs) and hyper-luminous supernovae can indeed be powered by
recently formed millisecond magnetars \citep{metzger11,bucciantini12}, although no observational evidence of the existence of such fast spinning
and strongly magnetized neutron stars have been collected thus far.
  
Besides the fast spinning proto-neutron star, a further idea on the origin of these high magnetic fields is that they simply reflect the high
magnetic field of their progenitor stars. Magnetic flux conservation \citep{wolt64} implies that magnetars must then be the stellar remnants of
stars with internal magnetic fields of $B > 1$ kG, whereas normal radio pulsars must be the end products of less magnetic massive stars.

Recent theoretical studies showed that there is a wide spread in white dwarf progenitor magnetic fields \citep{wick05}, which, when extrapolated
to the more massive progenitors implies a similar wide spread in neutron stars progenitors \citep{fer06}. Hence, apparently it seems that a
fossil magnetic field might be the solution of the origin of such strongly magnetized neutron stars, without the need of invoking dynamo actions
on utterly fast spinning proto-neutron stars.


\begin{table}
\tiny
\centering
\begin{tabular}{@{}cccccc}
\hline
\bf{SNR} & \bf{Instrument} & \bf{ObsID} & \bf{Date} & \bf{Detector} & \bf{Exp. (s)}\\
\hline
Kes73 & XMM & 0013340101 & 2002-10-05 & PN & 6017\\
 & & & & MOS1 & 5773\\
 & & & & MOS2 & 5771\\
 & & 0013340201 & 2002-10-07 & PN & 6613\\
 & & & & MOS1 & 6372\\
 & & & & MOS2 & 6372\\
CTB 109 & XMM & 0057540101 & 2002-01-22 & PN & 12237\\
 & & & & MOS1 & 19027\\
 & & & & MOS2 & 19026\\
 & & 0057540201 & 2002-07-09 & PN & 14298\\
 & & & & MOS1 & 17679\\
 & & & & MOS2 & 17679\\
 & & 0057540301 & 2002-07-09 & PN & 14011\\
 & & & & MOS1 & 17379\\
 & & & & MOS2 & 17379\\
N49 & XMM & 0505310101 & 2007-11-10 & PN & 72172\\
Kes75 & Chandra & 748 & 2000-10-15 & ACIS-S & 37280\\
 & & 6686 & 2006-06-07 & ACIS-S & 54070\\
 & & 7337 & 2006-06-05 & ACIS-S & 17360\\
 & & 7338 & 2006-06-09 & ACIS-S & 39250\\
 & & 7339 & 2006-06-12 & ACIS-S & 44110\\
 \hline
 \hline
\end{tabular}
\caption{Observations used in this paper.}
\label{tab:obs}
\end{table}


However, this lead to the problem of the formation of such high B progenitor stars. The most common idea is that the magnetic field in the star
reflects the magnetic field of the cloud from which the star is formed. The best studied very massive stars (around $\sim$40 M$_{\odot}$) with a
directly measured magnetic field are $\theta$ Orion C and HD191612, with dipolar magnetic field of 1.1 kG and 1.5 kG, respectively
\citep{donati02,donati06}. Very interestingly, the magnetic fluxes of both these stars ($1.1 \times 10^{27}$ G~cm$^2$ for $\theta$ Orion C and
$7.5 \times 10^{27}$ G cm$^2$ for HD191612) are comparable to the flux of the highest field magnetar SGR 1806-20 ($5.7\times 10^{27}$ G cm$^2$;
\citealt{woods06}). Other high magnetic field stars are reported in \cite{ost11}.

Recent observations of the environment of some magnetars revealed strong evidence that these objects are formed from the explosion of very
massive progenitors (M $>$30 M$_{\odot}$). In particular: i) a shell of HI has been detected around \ee , and interpreted as ISM
displaced by the wind of a progenitor of 30--40\, M$_{\odot}$ \citep{gaensler05}; \sgrb\, and \sgra have been claimed to be hosted by very
young and massive star clusters, providing a limit on their progenitor mass of $>$ 50 M$_{\odot}$ \citep{fuchs99,figer05,davies09} and
$>$20 M$_{\odot}$ \citep{vrba00}, respectively. Finally, \cxo\ is a member of the massive cluster Westerlund 1 \citep{muno06,ritchie10}, with a
progenitor with mass estimated to be $>$ 40 M$_{\odot}$ (see also \citealt{clark14}).
  
\citet{vink06} have started the idea of studying the energetics of supernova remnants surrounding magnetar with the aim of disentangling a
possible energetic difference between those remnants and others surrounding normal pulsars. Their work did not find any clear evidence i.e. of an
additional energy released in the remnant possibly due to an excess of rotational energy at birth.

Following this study we decided to extend their work re-analyzing all available \xmm\, or \chandra\, data of all confirmed and bright SNRs
associated with a magnetar or with a high-B pulsar showing magnetar-like activity, and comparing in a coherent and comprehensive way all the
extracted properties of these SNRs with other remnants: in particular line ionization and X-ray luminosity. In section \S\,\ref{data} we report
on the data analysis and reduction of our observational sample, in \S \ref{results} the results of our analysis, and we discuss our findings in
\S \ref{discussion}.

\section{Data analysis and reduction}
\label{data}

In this work, our approach has tried to be as conservative and model independent as possible. In particular, our target sample has been chosen
such to include all confirmed associations (see the McGill catalog\footnote{http://www.physics.mcgill.ca$/~$pulsar/magnetar/main.html} for all
proposed associations), and among those, we chose only those supernova remnants bright enough, and with sufficiently good spectra, to perform a
detailed analysis and classification of their spectral lines. We analyze the X-ray spectral lines of four SNRs hosting a neutron star that
showed magnetar-like activity in its center: Kes 73, CTB 109, N 49 and Kes 75. We use for all targets the best available archival data: from
the \xmm telescope in the case of Kes 73, CTB 109 and N 49, and \chandra for Kes 75. The observations used are summarized in Table \ref{tab:obs}.
To compare coherently all the spectral lines and fluxes we observed for these remnants we have chosen to use an empirical spectral fitting for
all SNRs. We have modeled all spectra using one or two Bremsstrahlung models for the spectral continuum, plus Gaussian functions for each
detected spectral line. We added spectral lines one by one until the addition of a further line did not significantly improve the fit (by using
the F-test). This approach is totally empirical, with respect of using more detailed ionized plasma models, but ensures a coherent comparison
between different remnants. In Table \ref{plasmamodels}, we report also the results of our spectra modeled with ionized plasma models, for a
comparison with the literature.

\begin{figure*}
\centering
\includegraphics[width=6cm, height=6cm]{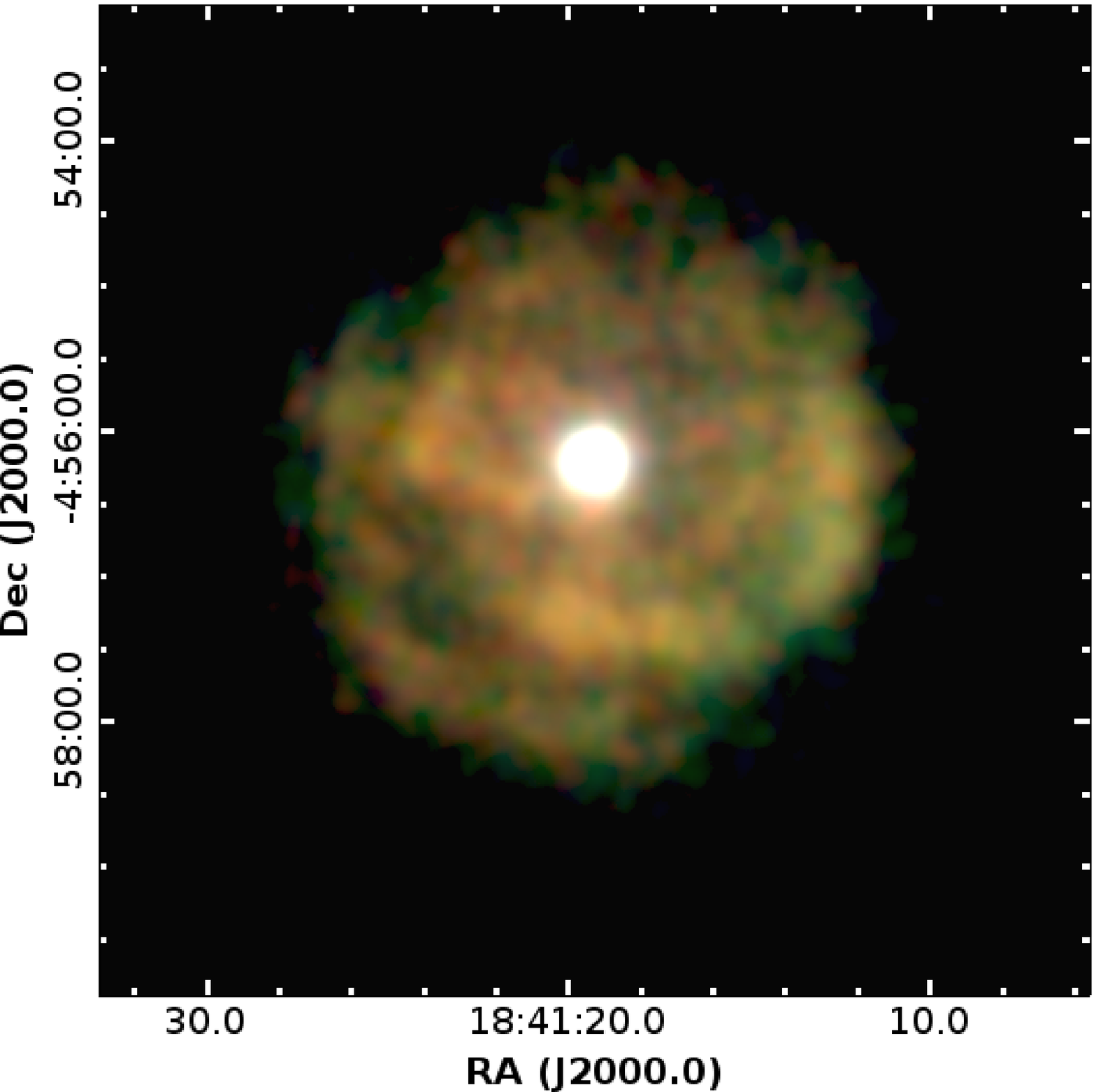}
\includegraphics[width=6cm, height=6cm]{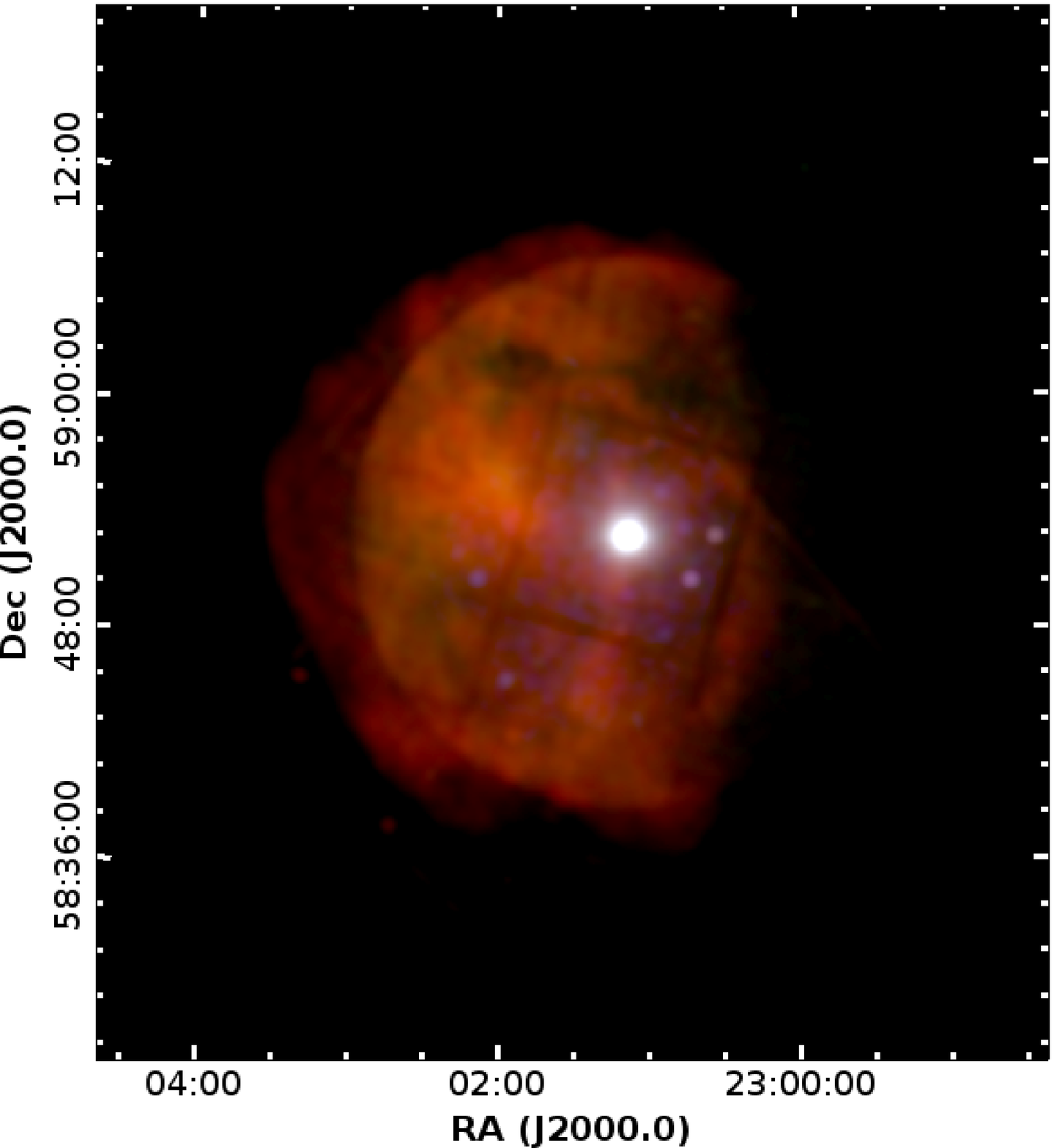}
\includegraphics[width=6cm, height=6cm]{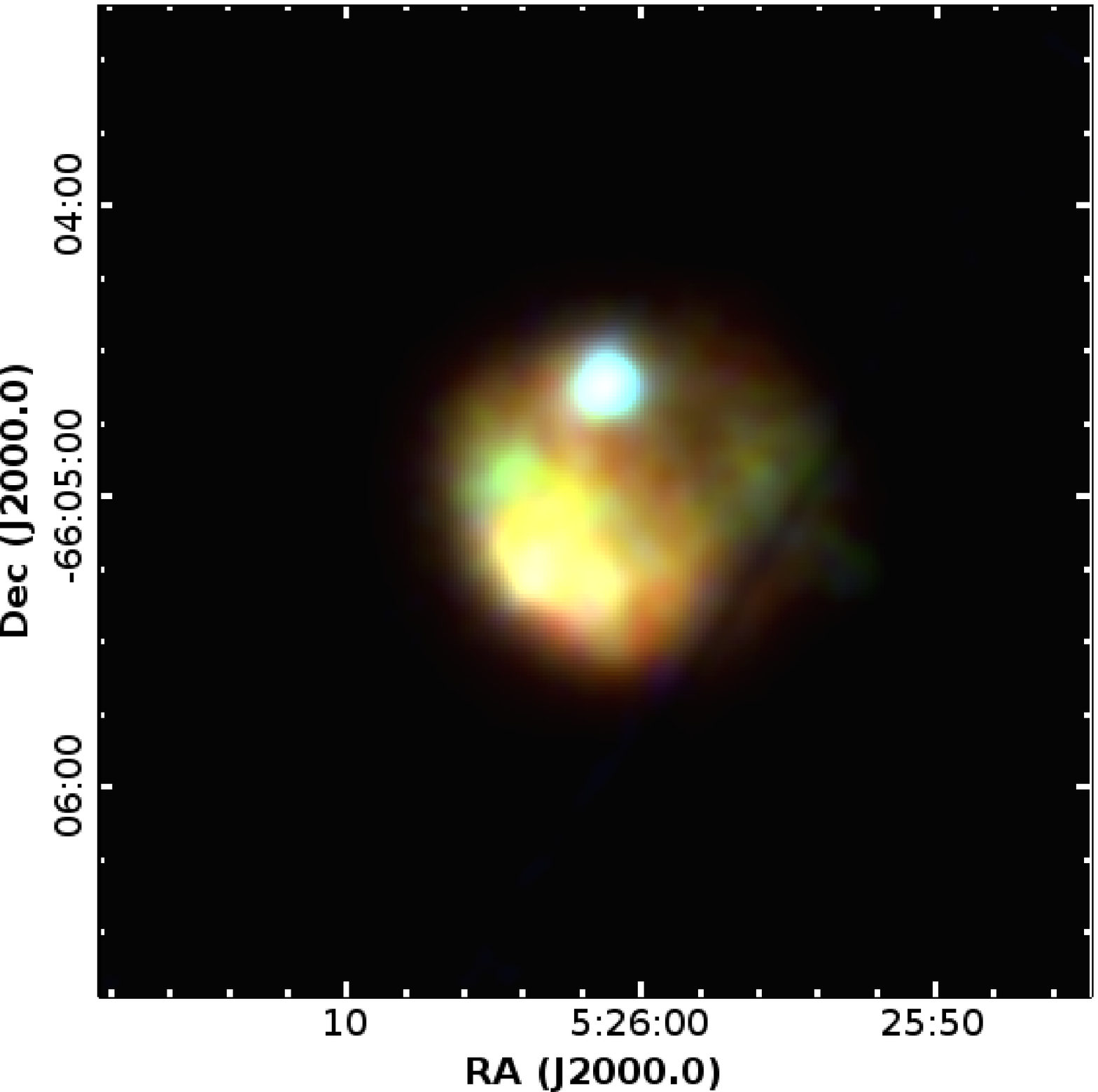}
\includegraphics[width=6cm, height=6cm]{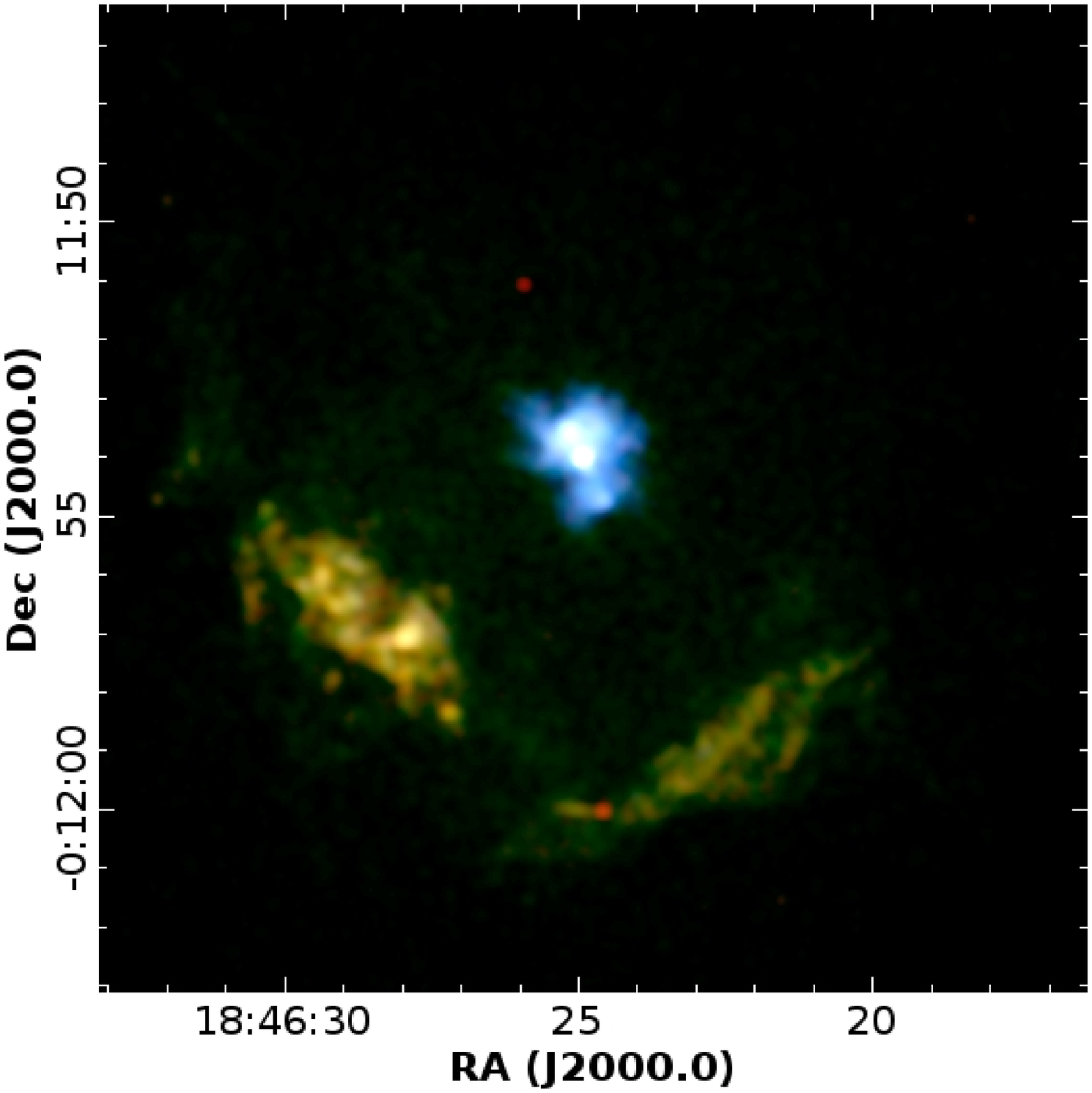}
\caption{Combined color images of Kes 73 (top-left), CTB 109 (top-right), N49 (bottom-left) and Kes 75 (bottom-right).}
\label{fig:neb}
\end{figure*}


\begin{figure*}
\centering
\includegraphics[width=6cm, height=6cm]{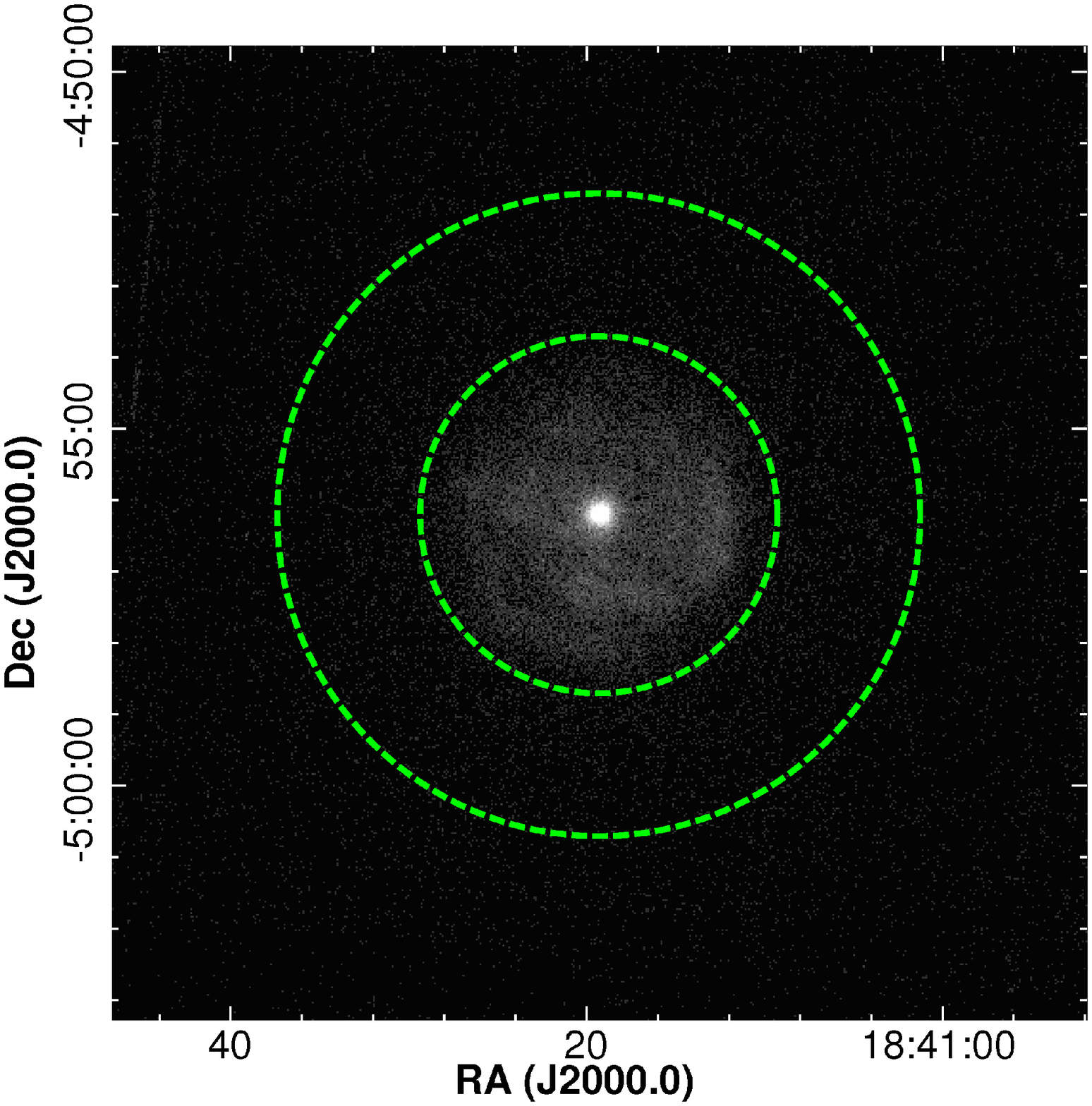}
\includegraphics[width=6cm, height=6cm]{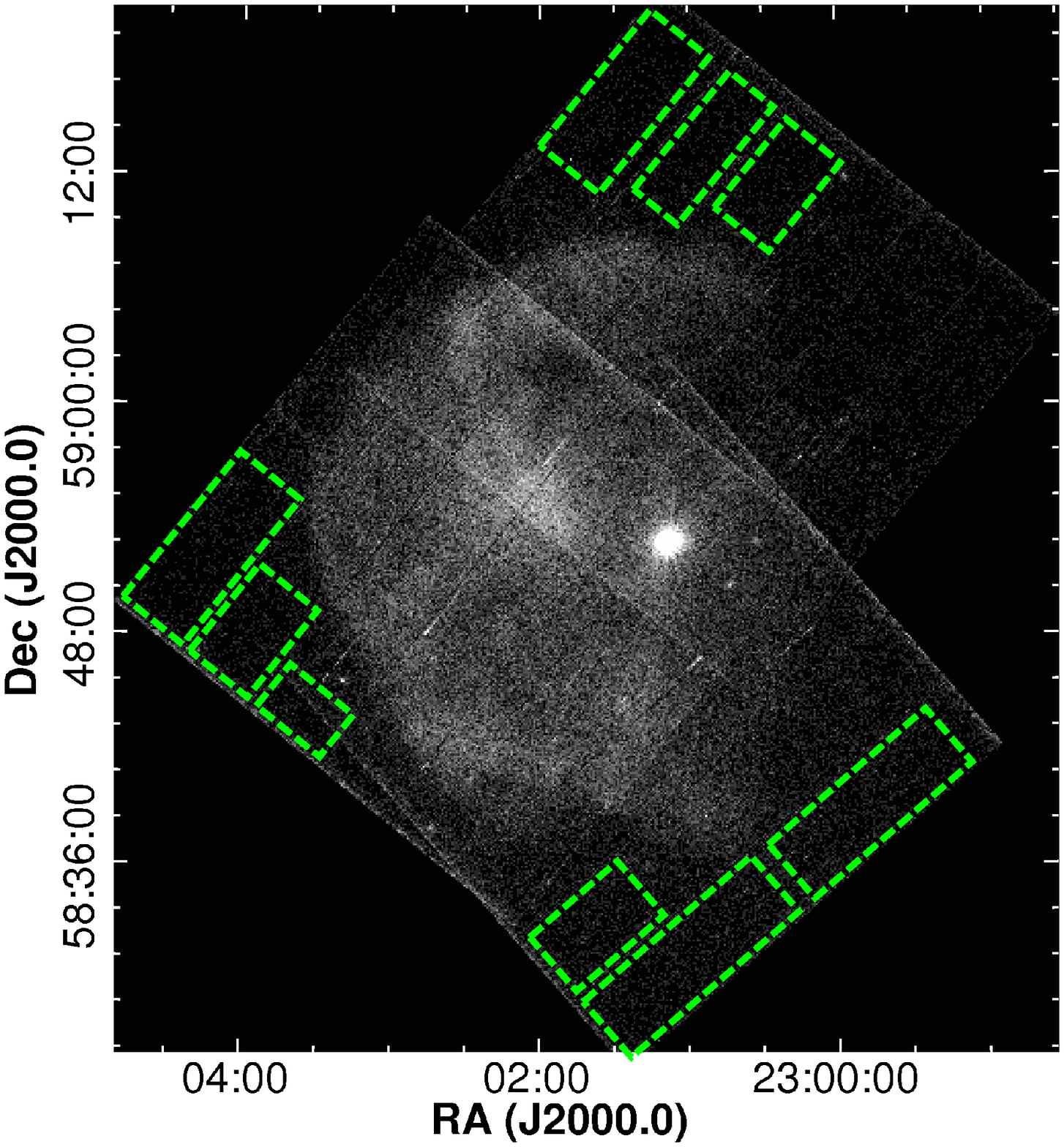}
\includegraphics[width=6cm, height=6cm]{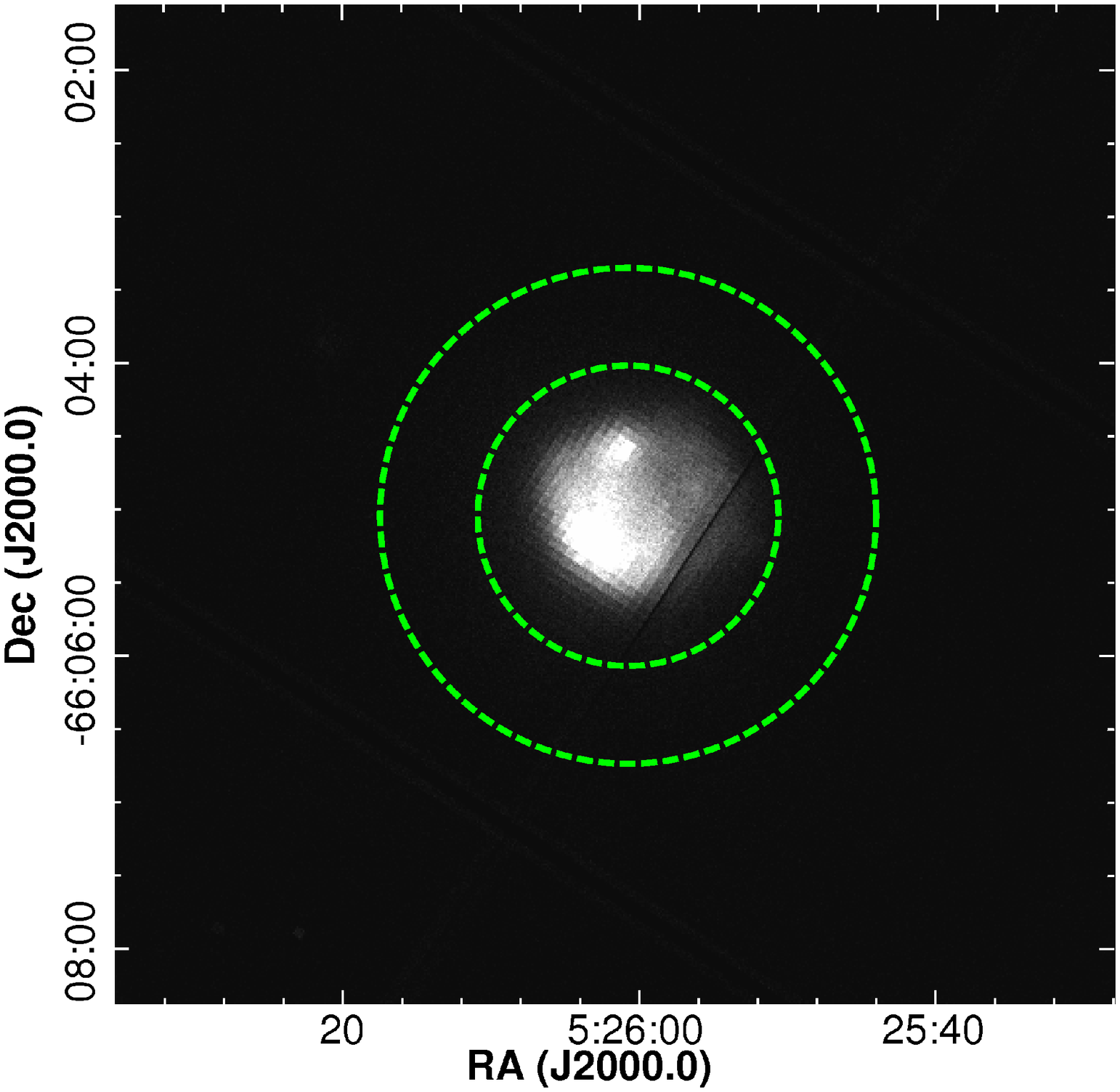}
\includegraphics[width=6cm, height=6cm]{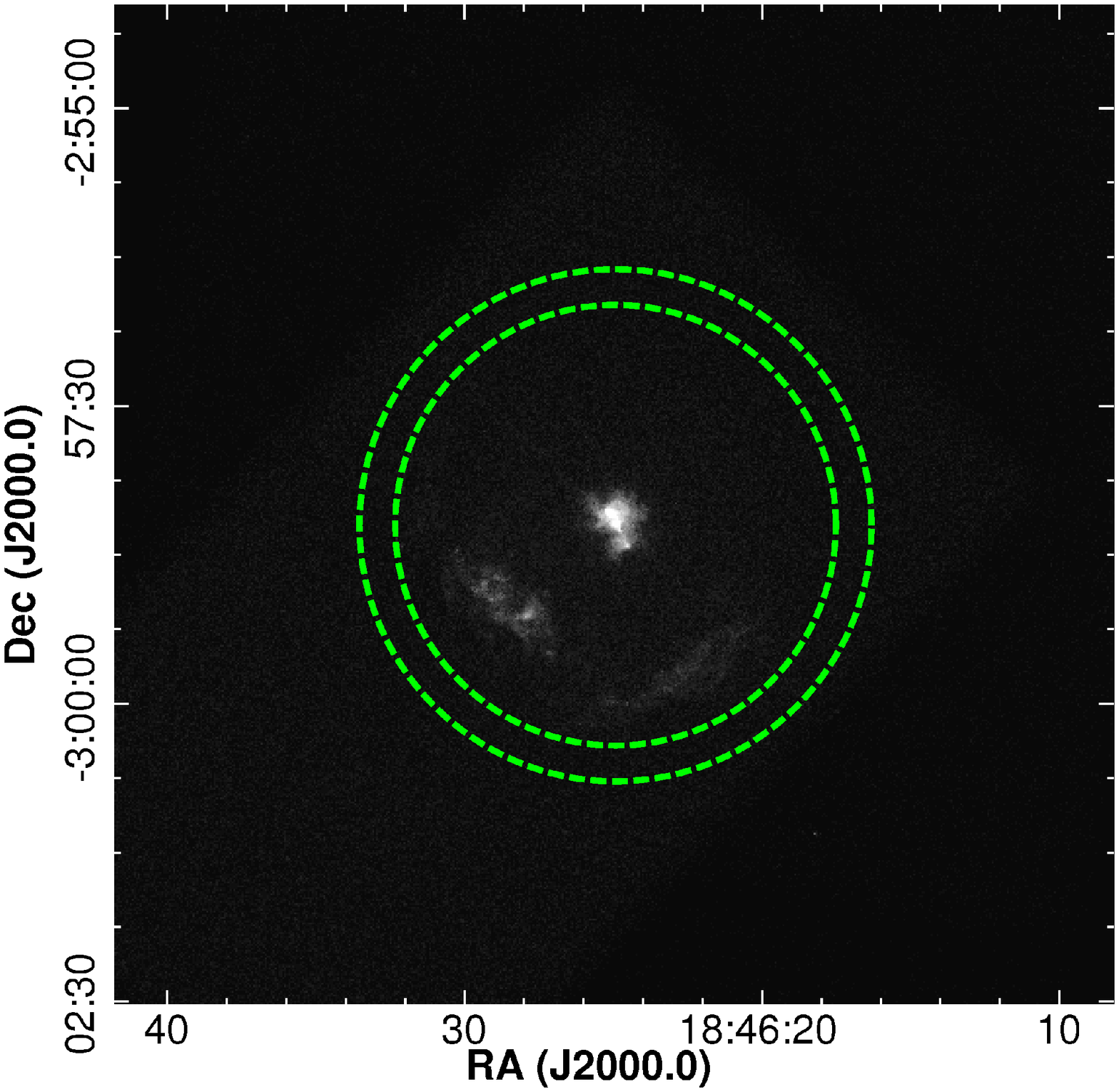}
\caption{Map of the backgrounds used in the spectrum analysis. The order of the images is the same as in figure~\ref{fig:neb}.}
\label{fig:bg}
\end{figure*}


\subsection{\xmm\ data}

We use images in full-frame mode obtained from the European Photon Imaging Camera (EPIC) PN \citep{struder01} and MOS \citep{turner01}. The
spectra of these images are fitted simultaneously in order to obtain the spectrum with the maximum possible number of counts. We used the
specific software for {\it XMM-Newton} data, Science Analysis System (SAS) v13.5.0 with the latest calibration files. To clean images of solar
flares, we used the SAS tool {\it tabtigen} to choose the good time intervals and extract them and the spectra with {\it evselect}. Source and
background spectra were extracted from each single image with pattern $\le$ 4 for PN images and pattern $\le$ 12 for MOS. The spectra and the
backgrounds corresponding to the same regions and the same detector were merged using the FTOOLS routine {\it mathpha} and we compute the
mean of the response matrices (RMF) and the ancillary files (ARF) weighted by the exposure time using the tools {\it addrmf} and {\it addarf}
(this means, that we keep PN, MOS1 and MOS2 data separately and we merge the spectra when they come from the same detector). Finally, we
binned the spectra demanding a minimum of 25 counts per bin to allow the use of $\chi^2$-statistics. 

We analyze the spectrum of each nebula considering its entire extension. For Kes 73, the nebula is completely covered in the EPIC PN, MOS 1 and
MOS 2 detectors and we consider all of them in the analysis. In the case of CTB 109, the SNR is too large to be included entirely in a single
pointing. The images with the \xmm data ID: 0057540101, 0057540201 and 0057540301 correspond to south, north and east pointings of the
remnant. We computed the spectra of each pointing, also considered the EPIC PN, MOS 1 and MOS 2 cameras. For N 49, the exposure time of the MOS detectors is very low in
comparison with PN. For this reason, we did not use the MOS data to avoid statistical noise in the data.

\subsection{\chandra\ data}

In the case of Kes 75, the best available observations were performed with \chandra using the Advanced CCD Imaging Spectrometer (ACIS). The ID
numbers of the data used are in Table \ref{tab:obs}. We used the standard reduction software for \chandra, the Chandra Interactive Analysis of
Observations (CIAO) v4.5. The spectra and the backgrounds were extracted using the routine {\it specextract} and the RMFs and ARFs using
{\it mkacisrmf} and {\it mkwarf} respectively. Finally, we combine the spectra demanding a minimum of 25 counts per energy bin using
{\it combine\_spectra}.

\section{Spectral analysis and Results}
\label{results}

\begin{figure*}
\centering
\includegraphics[scale=0.4]{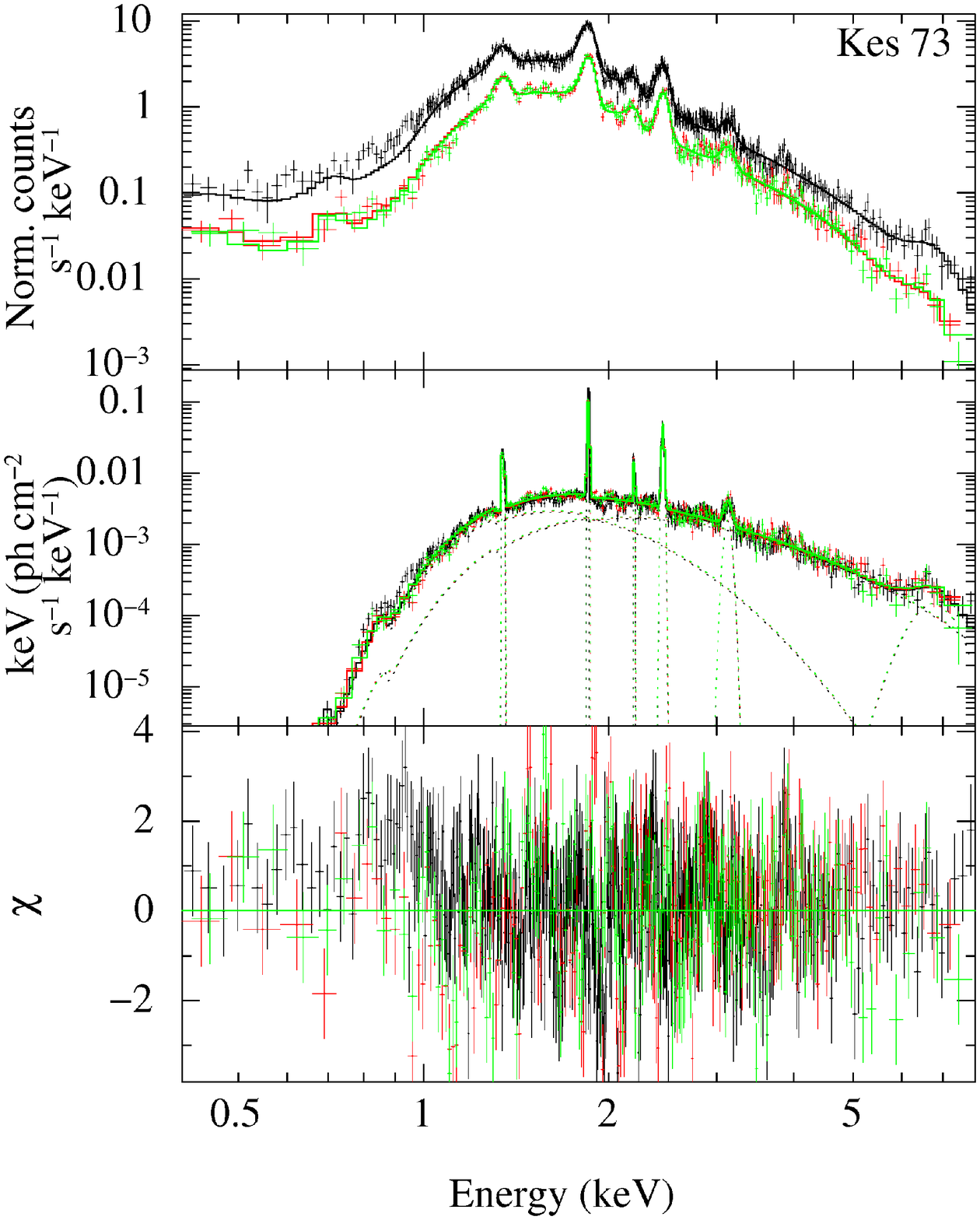}
\includegraphics[scale=0.4]{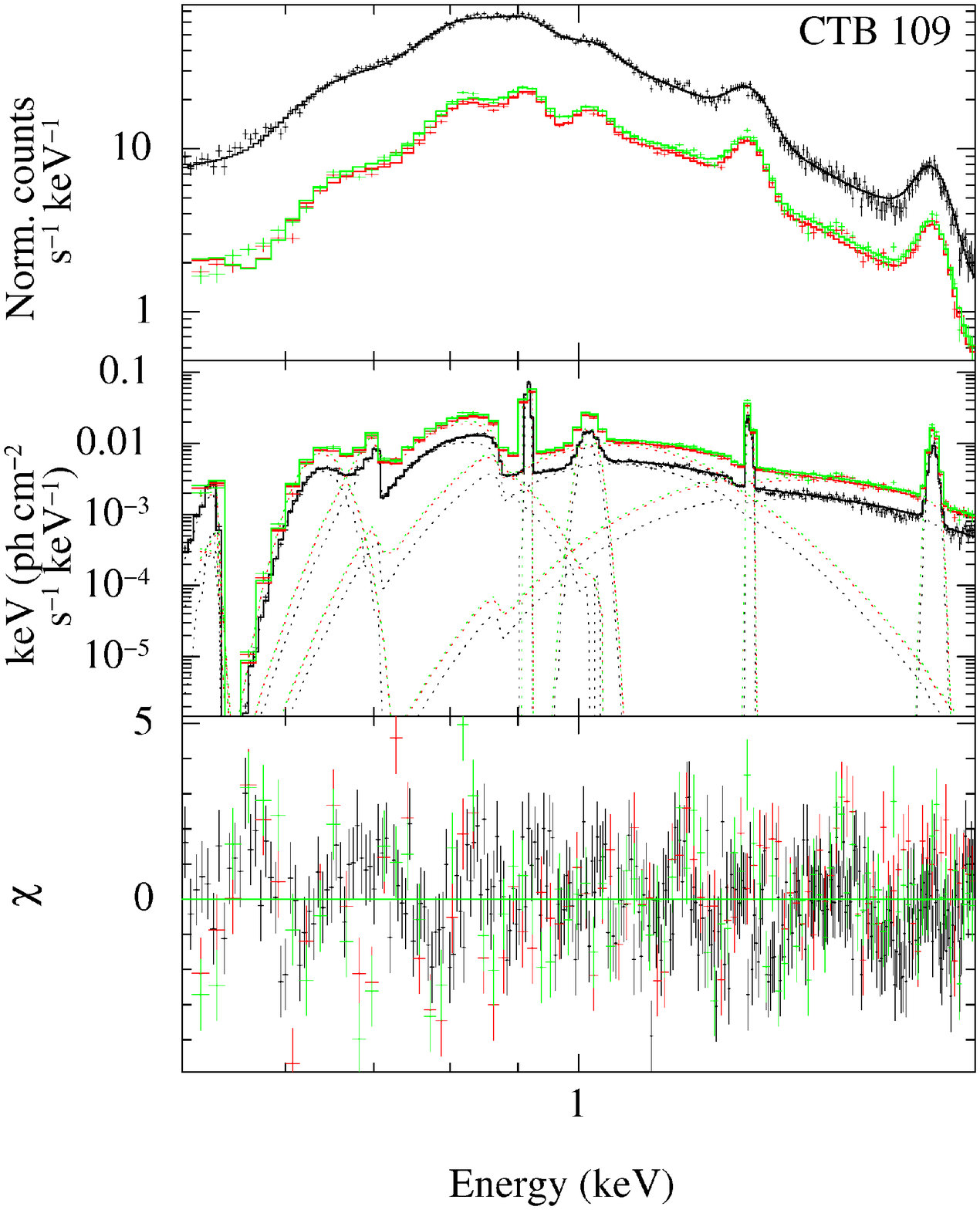}\\
\includegraphics[scale=0.4]{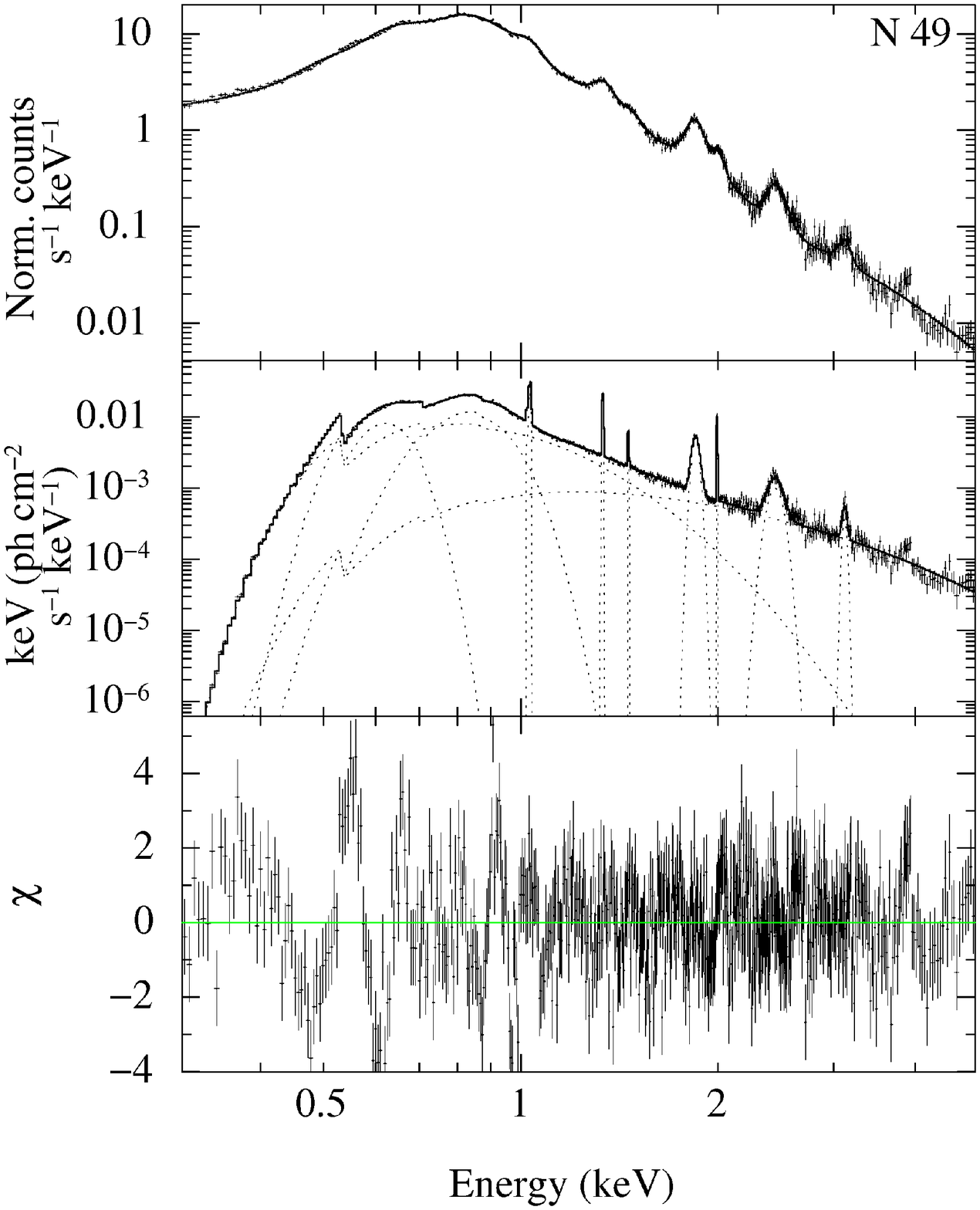}
\includegraphics[scale=0.4]{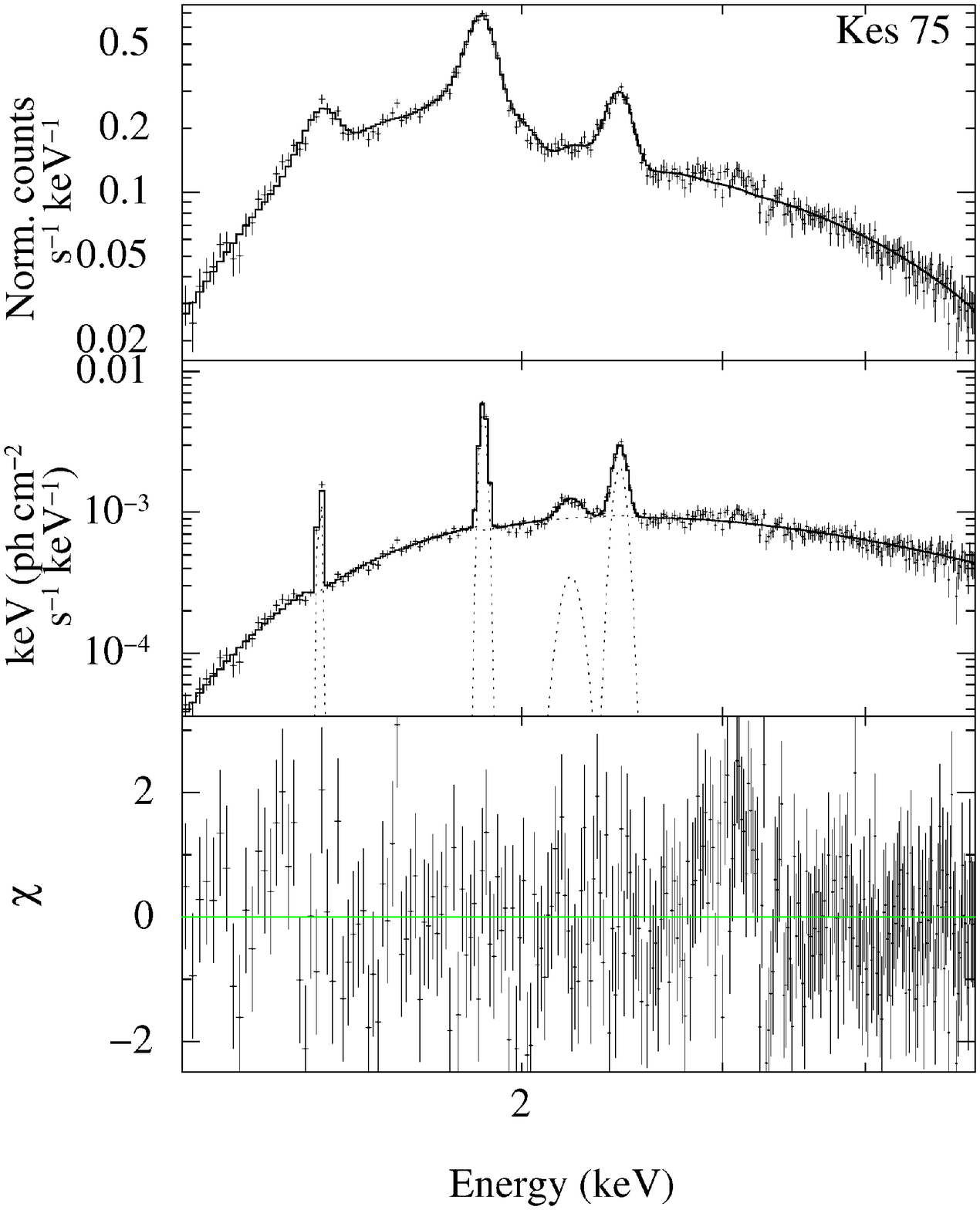}
\caption{Spectra obtained for the Kes 73, CTB 109, N 49 \& Kes 75. We used the EPIC PN (in black), MOS 1 (in red) and MOS 2 (in green) data
simultaneously to fit the models.}
\label{fig:spec}
\end{figure*}

We report the fitted spectra in figure \ref{fig:spec}, while reporting the best fitting models and relative parameters in Tables \ref{tab1_snr}
and \ref{tab2_snr}. For the spectral analysis, we used the program {\em XSPEC} \citep{arnaud96} v12.8.1 from the package HEASOFT v6.15. As
anticipated above, we have used for all SNRs a spectral model comprised of photo-electric absorption ({\tt phabs}), one or two Bremsstrahlung
models ({\tt brems}), plus a series of Gaussian functions to model the emission lines. Even if more physical ionized plasma models such a
{\tt vnei}, {\tt vshock} or {\tt vpshock} could be used to fit those SNRs: e. g., \citet{kumar14} for Kes 73, \citet{sasaki04,sasaki13} for CTB
109, \citet{park12} for N 49 and \citet{temim12} for Kes 75; we prefer to use a more empirical approach to compare coherently the emission lines
and luminosities of those objects, which is the aim of our work. Below we summarize for each studied remnant our results in the context of the
general properties of the SNR.

In Figure \ref{fig:bg} we show the background regions we have chosen for this analysis. We have tried several different regions finding consistent results. During the spectral analysis we checked that subtracting the background spectra or fitting it separately from the remnant spectra and subtracting its best fitting model, gave consistent results.

\subsection{Kes 73}
\label{sec:res_kes73}

Kes 73 (also known as G27.4+0.0) is a shell-type SNR. Its dimensions are about $4.7' \times 4.5'$ and it is located between 7.5 and 9.8 kpc
\citep{tian08b}. The central source is the magnetar 1E 1841$-$045 discovered as a compact X-ray source with the Einstein Observatory
\citep{kriss85}, and confirmed as a magnetar in \citet{vasisht97,gott99b}. The period of the magnetar is 11.78 s and its period derivative is
4.47 $\times 10^{-11}$ s s$^{-1}$. The resulting dipolar magnetic field is 7.3 $\times 10^{14}$ G, the spin-down luminosity is 1.1 $\times 10^{33}$
erg s$^{-1}$ and the characteristic age is 4180 yr. The age of the SNR shell is estimated around 1300 yr \citep{vink06}, which is consistent with
the age between 750 and 2100 yr estimated by \citet{kumar14}. Kes 73 has been also observed by {\em ROSAT} \citep{helfand94}, {\em ASCA} \citep{gott97},
{\em Chandra} \citep{lopez11} and {\em Suzaku} \citep{sezer10}.

Kes 73 shows a quite spherical structure with 1E 1841-045 in the center of the remnant (see Figure \ref{fig:neb}). In the
western part of the nebula (right-hand side of the images), we distinguish a shock ring which encloses the central source from west to east of the
image passing below the central source. Most of the flux is emitted between 1 and 3 keV. Finally, we analyzed the total spectrum of the nebula
excluding a circle of 40$"$ around the central source to exclude possible contamination from the central object. The background spectrum has been
extracted from a surrounding annular region shown in Figure \ref{fig:bg}, avoiding gaps between the CCDs to ensure good convergence of the
response matrices. The continuum spectrum has been fitted with two plasmas
with temperatures of 0.43 keV and 1.34 keV.  The absorption column density obtained is $N_{\rm H}=2 \times 10^{22}$ cm$^{-2}$. We detected 6
lines. The most prominent is the Fe XXV at 6.7 keV with an equivalent width (EW) of 1.89 keV. Other lines are Mg XI at 1.35 keV (EW=95 eV), Si
XIII at 1.85 keV (EW=0.37 keV), Si XIII at 2.19 keV (EW=46 eV), S XV at 2.45 keV (EW=0.38 keV) and Ar XVII at 3.13 keV (EW=0.12 keV).

\subsection{CTB 109}
\label{sec:res_ctb}

CTB 109 (also G109.2-1.0) was discovered in X-rays with the Einstein Observatory by \citet{greg80}, it is $30' \times 45'$ wide and the estimated
distance is about 3 kpc \citep{kothes02}. The central source is the magnetar 1E 2259+586 with a spin period of 6.98 s \citep{fahlman83} and
a period derivative of 4.83 $\times 10^{-13}$ \citep{iwasawa92}. The dipolar magnetic field is about 5.9 $\times 10^{13}$ G, the spin down power
is 5.6 $\times 10^{31}$ erg s$^{-1}$ and the characteristic age is 229 kyr. Despite the large characteristic age of the pulsar, the estimated
true age of the remnant is about 14 kyr \citep{sasaki13}. CTB 109 has been observed also in X-rays with {\em ASCA} \citep{rho98}, {\em BeppoSAX}
\citep{parmar98} and {\em ROSAT} \citep{hurford95,rho97}.

The spectrum covers the entire shell and combines the three observations detailed in Table \ref{tab:obs}. The background regions used are shown
in Figure \ref{fig:bg}. We observe that the main contribution to the flux is in the energy range between 0.5 and 2 keV. Some known X-ray sources
in the field of view have been excluded in our analysis. 


In this case we used two Bremsstrahlung models to fit the continuum, with temperatures of 0.07 keV and 0.20 keV. The measured absorption density
is $N_{\rm H}=$2.83 $\times 10^{22}$ cm$^{-2}$, and we detected 6 lines: N VII at 0.52 keV (EW=0.74 keV) and at 0.60 keV (EW=0.47 keV),
Ne IX at 0.91 keV (EW=0.15 keV), Ne X at 1.01 keV (EW=68 eV), Mg XI at 1.35 keV (EW=0.34 keV) and Si XIII at 1.86 keV (0.28 keV).

\subsection{N 49}
\label{sec:res_n49}

N49 (also SNR B0525-66.1) is a SNR located in the Large Magellanic Cloud (LMC). The associated central source is SGR 0526-66 with a period of
8.047 s \citep{mazets79} and a period derivative of 6.6 $\times 10^{-11}$ s s$^{-1}$ \citep{kulk03}. There is some uncertainty in the association
of SGR 0526-66 with N49 (see \citealt{gaensler01}). The inferred dipolar magnetic field is 7.3 $\times 10^{14}$ G, the spin-down luminosity is
4.9 $\times 10^{33}$ erg/s and the characteristic age is $\sim$2 kyr. The nebula is $1.5' \times 1.5'$, this means that assuming a distance of 50
kpc the diameter of N49 is $\sim$22 pc. \citet{park12} establish a Sedov age for the nebula of $\sim$4.8 kyr and a SN explosion energy of 1.8
$\times 10^{51}$ erg.

SGR 0526-66 is located in the north of the remnant. The brightest part of the nebula is in the southeast, coinciding with dense interstellar
clouds \citep{vancura92,banas97,park12}. This part of the remnant also has contributions between 3 and 10 keV, while the contribution of the rest
of the nebula is clearly negligible at this range. In Figure \ref{fig:neb}, we show a color image of N49. We analyze the total spectrum of the
nebula excluding a circle of 20$"$ around the central source to avoid its contribution to the spectrum.

The absorption of N49 has two components: one is related with the Galactic absorption and the other is the absorption produced by LMC. The Milky
Way photoelectric absorption towards N49 is fixed as $N_{\rm H}=6 \times 10^{20}$ cm$^{-2}$ \citep{dickey90,park12}. We include a second absorption component to take into account the absorption column density for LMC, where we use the abundances given by \citet{russell92,hughes98,park12}. We obtain an absorption column density
of $N_{\rm H}=0.7 \times 10^{22}$ cm$^{-2}$ for the LMC contribution. The continuum is represented by two Bremsstrahlung models with temperatures of 0.23 keV and 1.14
keV. In this case, we have detected 9 lines: O VII at 0.57 keV (EW=0.20 keV), O VIII/Fe XVIII at 0.77 keV (EW=0.34 keV), Ne X at 1.03 keV (EW=33
eV), Mg XI at 1.33 keV (EW=62 eV), Mg XII at 1.46 keV (EW=20 eV), Si XIII at 1.85 keV (EW=0.30 keV), Si XIV at 2.00 keV (EW=0.13 keV), S XV at
2.44 keV (EW=0.30 keV) and Ar XVII at 3.12 keV (EW=0.11 keV).

\begin{figure*}
\centering
\includegraphics[scale=0.32]{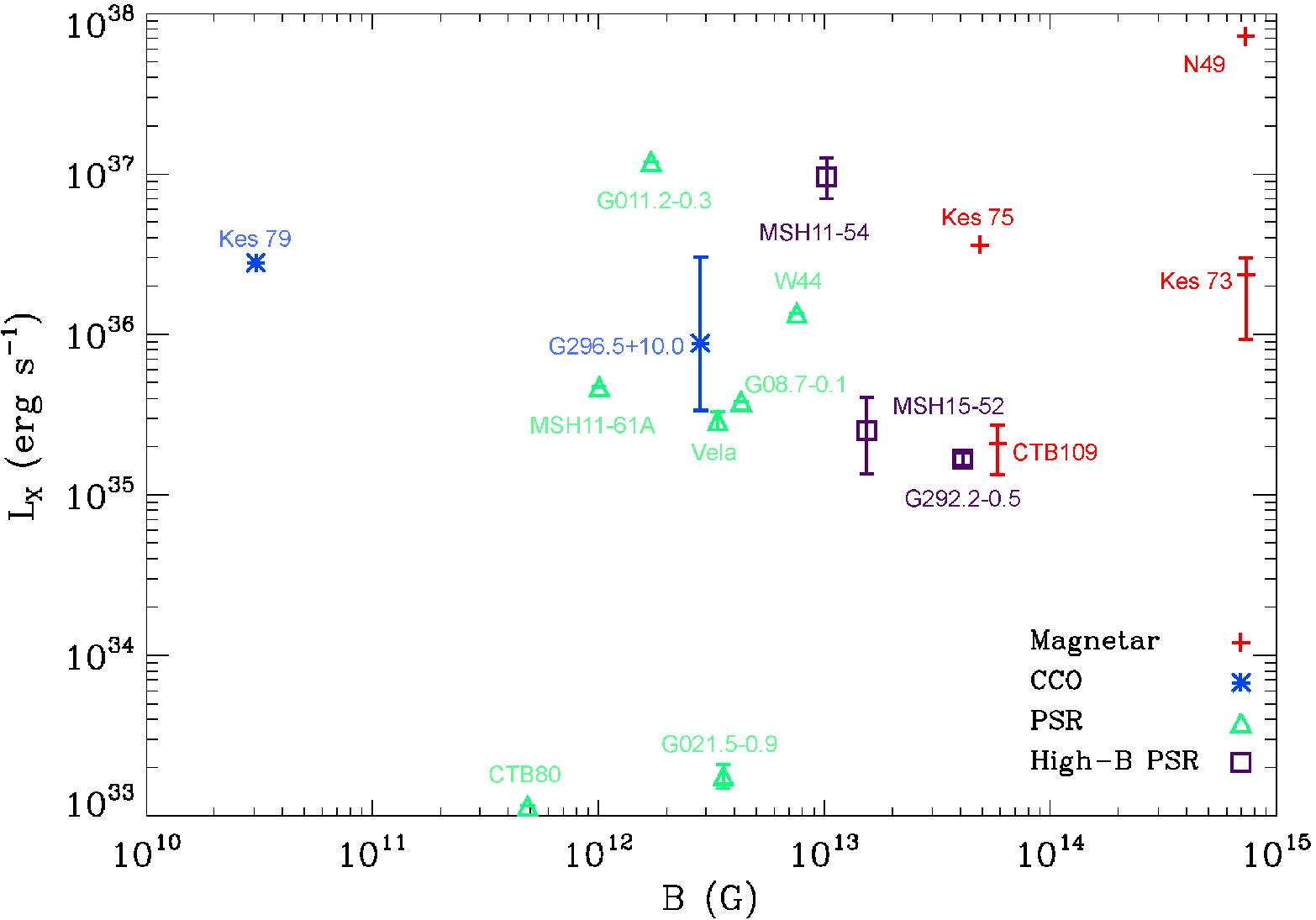}
\includegraphics[scale=0.32]{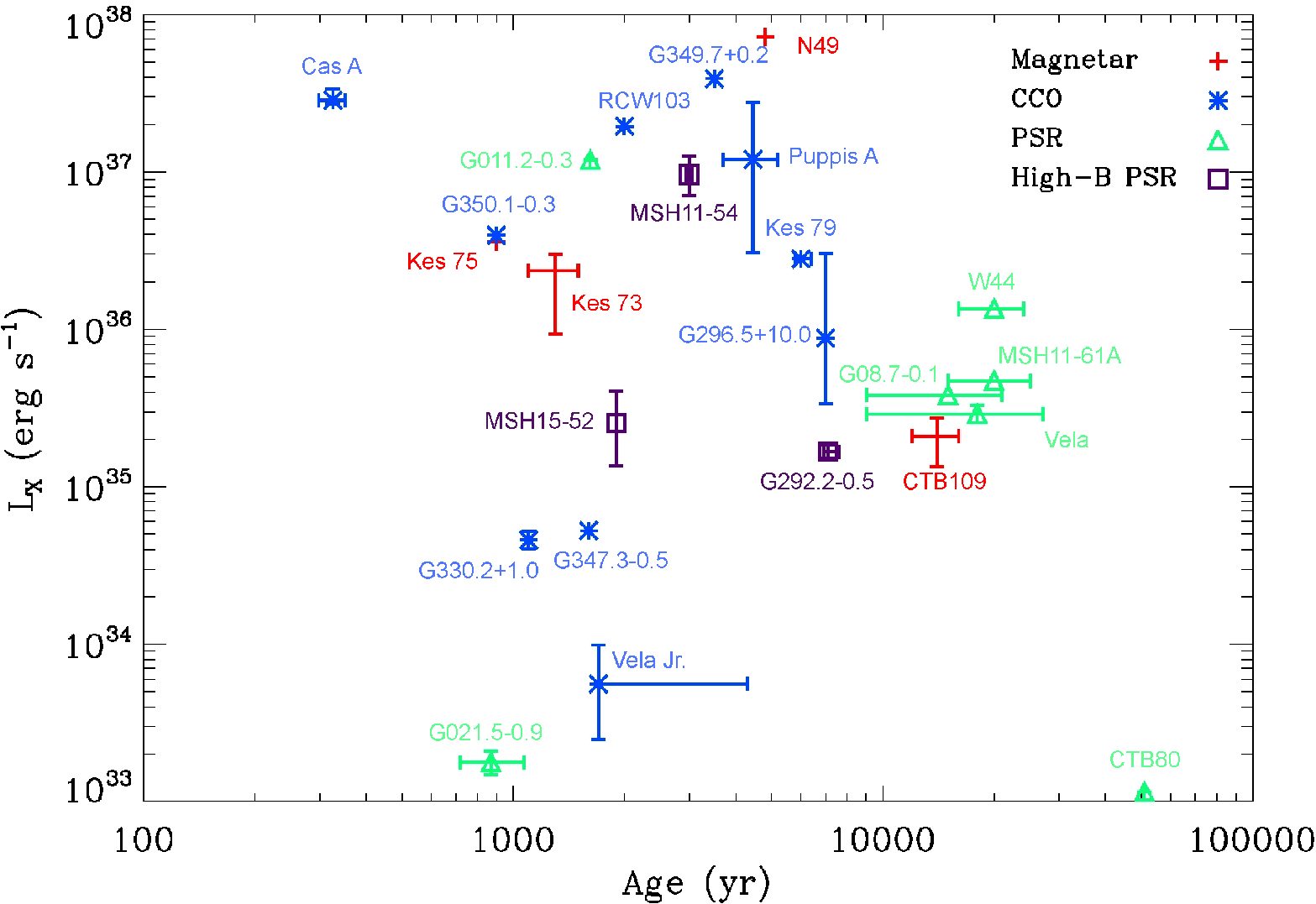}
\includegraphics[scale=0.32]{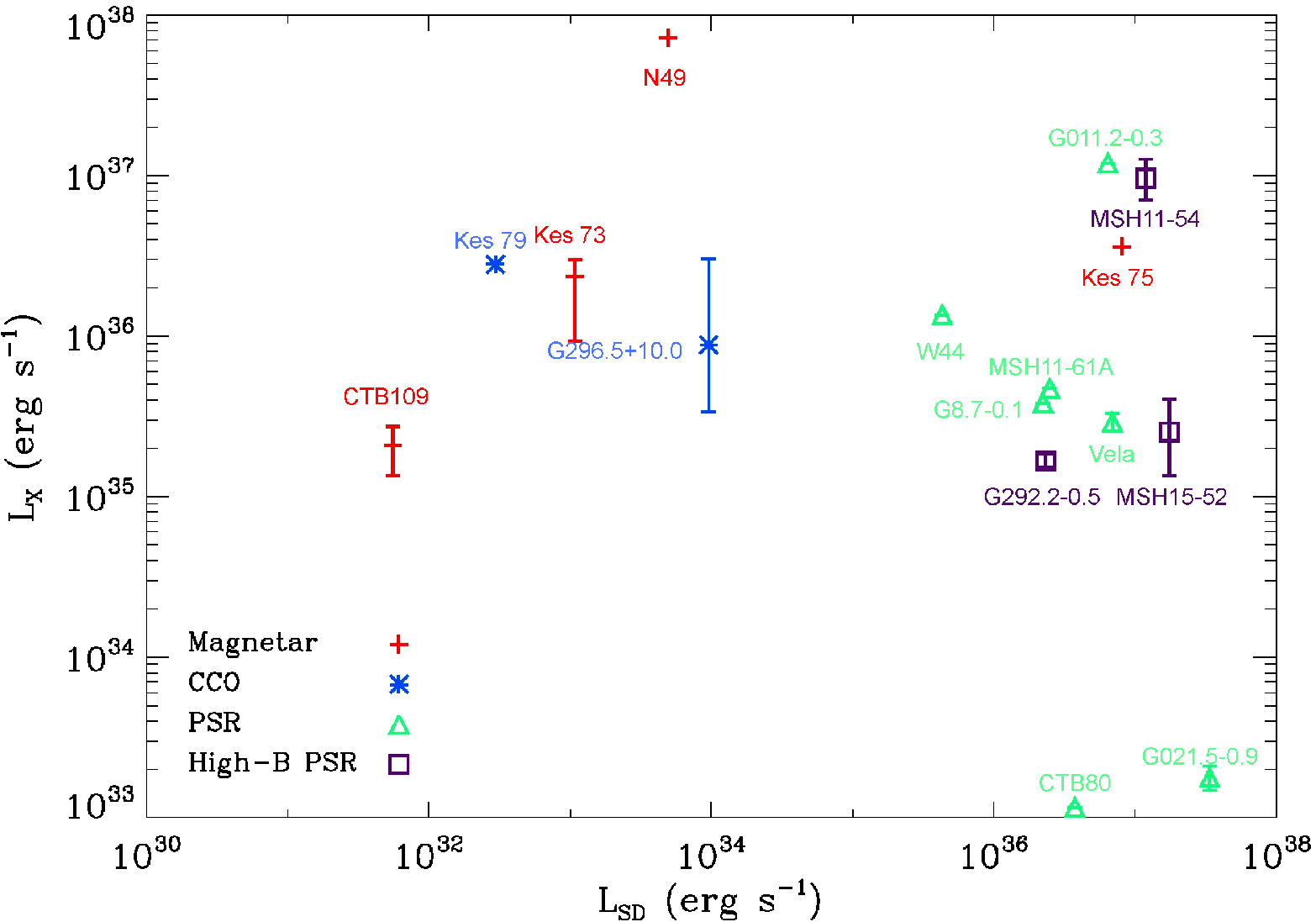}
\includegraphics[scale=0.32]{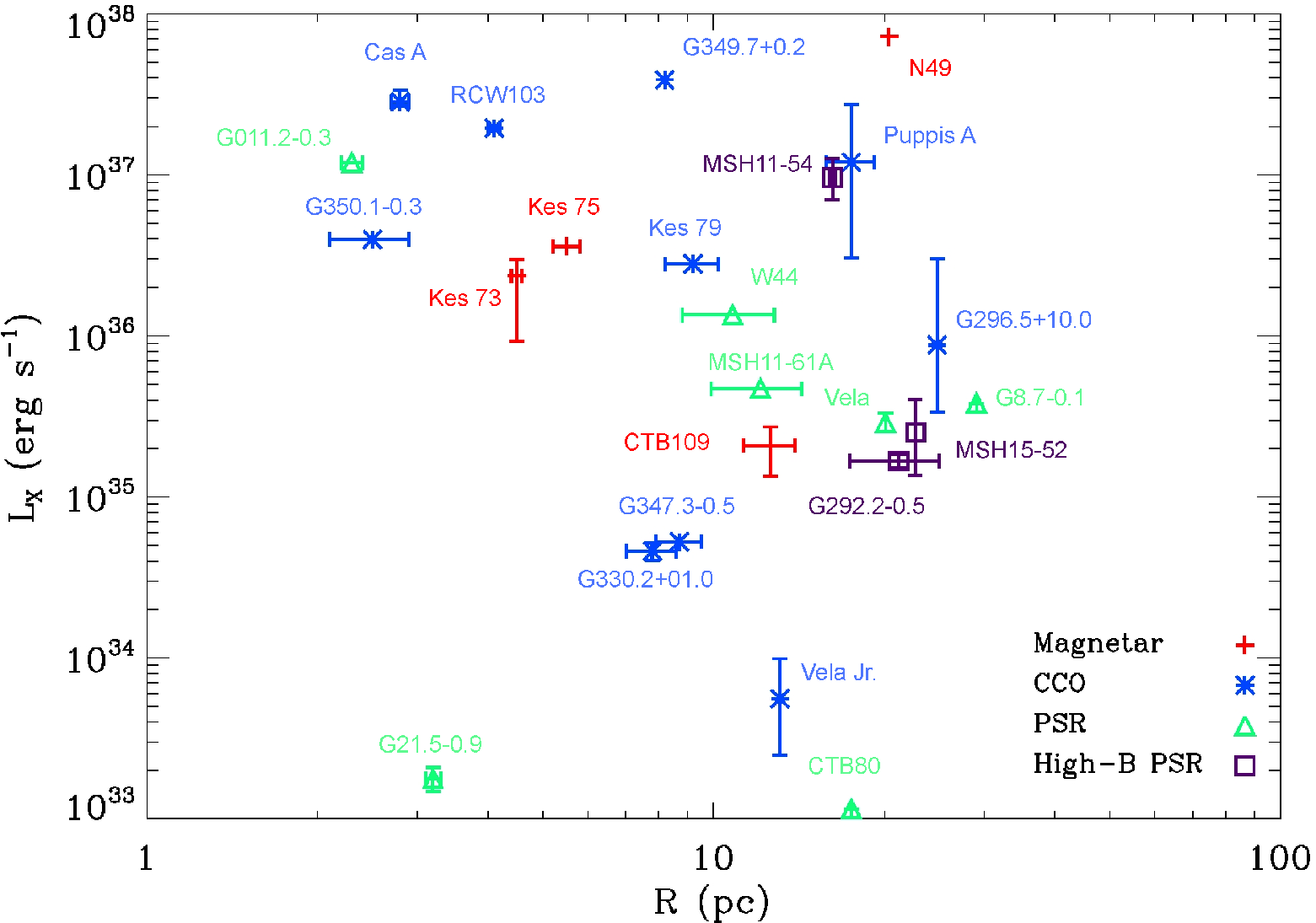}
\caption{X-ray luminosity of all observed, and securely associated, X-ray emitting SNRs containing a magnetar, a CCO, a high-B pulsar or a
normal pulsar, plotted versus magnetic-field (top-left), age (top-right), spin-down luminosity (bottom-left) and remnant radius (bottom-right).}
\label{xvsage_a}
\label{xvsage_r}
\label{xvsage_r}
\label{xvsrad_sp}
\end{figure*}

\subsection{Kes 75}

Kes 75 (G29.7-0.3) is a composite SNR. The X-ray emission of the partial shell is extended in two clouds in the southwest and southeast part
of the image (see Figure \ref{fig:neb}). It was observed firstly in X-rays by {\em Einstein} \citep{beck83} showing an incomplete shell of 3'
in extent. In the center of the nebula, there is a bright pulsar wind nebula (PWN), which was spatially resolved by the {\em Chandra} observation \citep{helf03,ng08},
and PSR J1846-0258 powers it. This pulsar was discovered using the {\it RXTE} telescope and localized within an arc minute of the remnant
using {\it ASCA} \citep{gott00}. The period of the pulsar is $\sim$326 ms and the period derivative $7.11 \times 10^{-12}$ s s$^{-1}$ (e. g.,
\citealt{liv11a}). This leads to a spin-down energy loss of $8.1 \times 10^{36}$ erg s$^{-1}$, a magnetic field of $4.9 \times 10^{13}$ G and a
characteristic age of 728 yr. \citet{liv06} estimated a braking index of 2.65$\pm$0.01. Despite its early classification as a typical rotational
powered pulsar, PSR J1846-0258 showed magnetar-like activity via short bursts and the outburst of its persistent emission
\citep{gavriil08,kumar08} enabling its classification as (at least sporadically) a magnetically powered pulsar. There is a big uncertainty in the distance
of this SNR in the literature \citep{casw75,milne79,mcbride08,beck84}. Most recent estimates give a distance between $\sim$5.1-7.5\,kpc based on
H I absorption observations \citep{leahy08}, and 10.6\,kpc using millimeter observations of CO lines from an adjacent molecular cloud
\citep{su09}. In our work, we adopt this value in order to compute the X-ray luminosity and the size of the SNR.

The spectrum of Kes 75 has been fitted using only one thermal Bremsstrahlung component with a temperature of 2.8 keV and an absorption column
density of $1.79 \times 10^{22}$ cm$^{-2}$. Four clear lines are resolved using Gaussians: Mg XI line at 1.33 keV (EW=84 eV), two Si XIII lines
at 1.85 (EW=0.23 keV) and 2.21 keV (EW=45 eV) and S XV at 2.44 keV (EW=0.18 keV).

\section{Discussion}
\label{discussion}

In this work we have re-analyzed in a coherent way the X-ray emission from SNRs around magnetars, and compared their emission lines and luminosities. The aim of this study was to search for any possible trend or significant difference in SNRs associated with different types of neutron stars. This work complements and extends the work by \citet{vink06}, providing a detailed description of the spectra for Kes 73, Kes 75, N 49 and CTB 109, and compares them directly with other remnants with similar spectroscopic X-ray studies. We also looked for any possible trend or significant difference in the ionization state and X-ray luminosity of SNRs associated with different types of neutron stars.

\subsection{Spectral lines comparison with other SNRs}

X-ray spectra of SNRs are usually fit with plasma models (see also Table \ref{plasmamodels}). In this work we proceed to fit the spectra of Kes 73, CTB 109, N 49 and Kes 75 using a thermal Bremsstrahlung model for the continuum emission and Gaussians for the lines. Our main aim is to have an estimate
of line centroid energy, to identify it properly. We have then used the simplest continuum model to reduce the free parameters of the fit\footnote{Note that in the 0.5-1\,keV the detection of spectral lines are dependent on absorption model we adopted.}. One
could expect that the excess of rotational energy released by the magnetar during the alpha-dynamo process could be stored in the ionization
level of the lines present in the spectrum. If the energy release is higher than in a normal SNR, heavy elements such as silicon (Si),
sulfur (S), argon (Ar), calcium (Ca) or iron (Fe) could be systematically at a higher state of ionization. In Table \ref{tab1_snr}, we collected
all SNRs with detailed spectroscopic studies in the literature and we see that the typical elements detected are O VII, O VIII, Ne IX, Ne X, Mg
XI, Mg XII, Si XIII, Si XIV, S XV, S XVI, Ar XVII, Ca XIX and Fe XXV. The only lines detected in all four of the spectra are the Mg XI line at
1.33 keV and Si XIII at 1.85 keV. For comparison, we also fitted the spectra of the SNRs using a {\tt vnei} model (e. g., \citealt{borkowski01}).
The results are summarized in the Table \ref{plasmamodels}. We have added a thermal Bremsstrahlung component in some cases. The temperature of
the {\tt vnei} plasma is always higher than for the thermal Bremsstrahlung, with the exception of N49 in which the temperature for {\tt vnei} is
0.17 keV (0.99 keV for Bremsstrahlung). The abundances obtained in both models show similar tendencies. For Kes 73 and N 49, the abundances of Si
and S are quite above the solar ones. CTB 109 shows low abundances with respect to the solar ones for O, Ne, Mg, Si and Fe. Due to the complexity
of the N 49 spectrum, some lines have not been reproduced well by the plasma models and we have added them using gaussian profiles to improve the
fit. In summary, our spectroscopic X-ray analysis of these sources shows compatible results with other non-magnetar SNRs already reported in literature.

\subsection{Comparison with other SNRs}

In Figure \ref{xvsage_a} we have collected from the literature the X-ray luminosities from 0.5 to 10 keV of all observed SNRs brighter than
$\sim10^{33}$ erg s$^{-1}$, with an age lower than 100 kyr and having a confirmed association with a central source. For these remnants, we obtain
the age, distance, approximate radius, magnetic field and spin-down luminosity of the central source (whenever possible) from the literature. All this
information is summarized in Table \ref{snrdata}. We have plotted the SNRs luminosities (excluding the contribution of the central neutron star
luminosity) as a function of the SNR age and dimension (although note that the latter parameter is highly dependent on the environment of each
remnant). For those remnants having a central neutron star with measured rotational properties, we plot the SNR luminosity as a function of the
pulsar surface dipolar magnetic field at the equator ($B=3.2\times10^{19}\sqrt{P\dot{P}}$ G), and the pulsar spin down luminosity
($L_{sd}= 3.9\times10^{46}\dot{P}/P^{3}$ erg/s; always assuming the neutron star moment of inertia $I=10^{45}$ g cm$^2$), and where $P$ is the
pulsar rotation period in seconds and $\dot{P}$ its first derivative.

\begin{figure}
\centering
\includegraphics[scale=0.8]{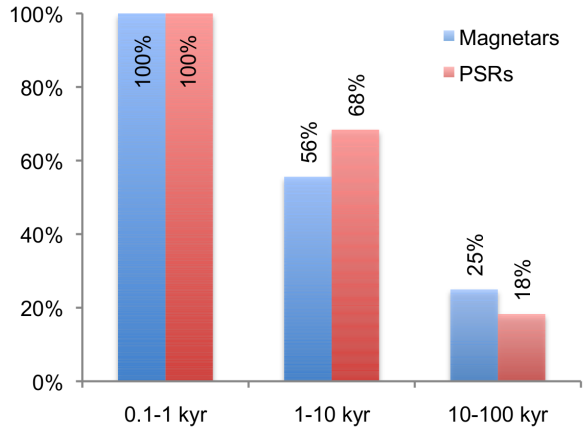}
\caption{Percentage of pulsars and magnetars having a detected SNR as a function of the age.}
\label{fig:percentage}
\end{figure}

In order to search for any correlations in the SNRs and pulsars characteristics (see Figure \ref{xvsage_a}), we run a Spearman test.
We searched for correlations between the X-ray luminosity and other features of the sources of our sample, such as dimension of the remnant, age,
surface magnetic field strength and spin down power of the associated pulsar. To this end, we employed a Spearman rank correlation test, and
evaluated the significance of the value of the coefficient of correlation $r$ obtained, by computing $t=r\sqrt{(N-2)/(1-r^2)}$, which is
distributed approximately as Student's distribution with $N-2$ degrees of freedom, where $N$ is the number of couples considered. The results we
obtained are listed in Table \ref{spearman}; no correlation is found at a significance level larger than 99\% , or any significant difference
in luminosity between SNRs surrounding magnetars and those around other classes of isolated neutron stars.

\begin{table}
\centering
\begin{tabular}{@{}lccc}
\hline
Parameters & \bf{r} & \bf{N} & \bf{p}\\
\hline
$L_X$ vs. age & -0.158 & 24 & 0.46\\
$L_X$ vs. radius & -0.245 & 24 & 0.25\\
$L_X$ vs. $B$ & 0.271 & 16 & 0.31\\
$L_X$ vs. $L_{sd}$ & -0.309 & 16 & 0.25\\
\hline
\end{tabular}
\caption{Spearman correlation coefficient (r), number of couples considered (N) and probability that the two samples are not correlated (p)
evaluated by comparing the X-ray luminosity of the sources of our sample with the age, radius, surface magnetic field strength and spin down
luminosity.}
\label{spearman}
\end{table}

We have also been looking at the number of pulsars having detected SNRs as a function of age,
and compared it to the magnetar case. We caution, however, that there are several systematic effects in this comparison (different detection
wavebands, distance, low number of magnetars in comparison with pulsars, etc.), but we were mostly interested in looking for a general trend. In Figure
\ref{fig:percentage} we plot the result of this comparison, where we can see how on average (with all the due caveats) for a
similar age, pulsars and magnetars seem to show a similar probability to have a detected SNR.

\section{Conclusions}

We have reported on the re-analysis of the X-ray emission of SNRs surrounding magnetars, using an empirical modeling of their spectrum with a
Bremsstrahlung continuum plus several emission lines modeled by Gaussian functions. Our analysis, and the comparison of the emission of those
remnants with other bright SNR surrounding normal pulsars suggest the following conclusions:

\begin{itemize}

\item We find no evidence of generally enhanced ionization states in the elements observed in magnetars' SNRs compared to remnants observed
around lower magnetic pulsars.

\item No significant correlation is observed between the SNRs X-ray luminosities and the pulsar magnetic fields.

\item We show evidence that the percentage of magnetars and pulsars hosted in a detectable SNR are very similar, at a similar age.

\end{itemize}

Our findings do not support the claim of magnetars being formed via more energetic supernovae, or having a large rotational energy budget at
birth that is released in the surrounding medium in the first phases of the magnetar  formation. However, we note that although we do not
find any hint in the SNRs to support such an idea, we cannot exclude that: 1) most of the rotational energy has been emitted via neutrinos or
gravitational waves, hence with no interaction with the remnant; or 2) we are restricted to a very small sample, and with larger statistics
some correlation might be observed in the future.

\vspace{1cm}

This work was supported by grants AYA2012-39303, SGR2009-811, iLINK 2011-0303 and the NewCOMPSTAR MP1304 COST Action. NR is supported by a Ram\'on y Cajal fellowship and by an NWO Vidi Award. AP is supported by a Juan de la Cierva fellowship. We are indebted to Samar Safi-Harb and Harsha Kumar for providing the
{\em Chandra} data on Kes\,75, and for useful comments. We also thank Manami Sasaki and the referee for comments and suggestions that improved the manuscript.

\newpage

\begin{table*}
\scriptsize
\begin{center}
\begin{tabular}{lllll}
\hline
\bf{Parameter} & {\bf Kes 73} & {\bf CTB 109} & {\bf N 49}$^{\dagger}$ & {\bf Kes 75}\\
\hline
$N_H$ ($10^{22} cm^{-2}$) & $2.00^{-0.02}_{+0.01}$ & $2.83^{-0.06}_{+0.10}$ & $0.698^{-0.024}_{+0.006}$ & $1.79^{-0.05}_{+0.06}$\\
$kT_1$ (keV) & $0.43^{-0.05}_{+0.02}$ & $0.065^{-0.002}_{+0.001}$ & $0.230^{-0.003}_{+0.004}$ & $2.8^{-0.1}_{+0.2}$\\
$N_1^{brems}$ (Norm. counts s$^{-1}$) & $0.36^{-0.02}_{+0.15}$ & $9^{-1}_{+14} \times 10^6$ & $0.512^{-0.007}_{+0.067}$ & $(4.5^{-0.3}_{+0.2}) \times 10^{-3}$\\
$kT_2$ (keV) & $1.34^{-0.01}_{+0.01}$ & $0.20^{-0.02}_{+0.03}$ & $1.14^{-0.01}_{+0.04}$ & -\\
$N_2^{brems}$ (Norm. counts s$^{-1}$) & $(2.47^{-0.06}_{+0.41}) \times 10^{-2}$ & $18^{-4}_{+9}$ & $(3.5^{-0.15}_{+0.08}) \times 10^{-3}$ & -\\
\hline
{\bf N VII (3,4 $\rightarrow$ 1)} & & & & \\
$E$ (keV) & - & $0.515^{-0.008}_{+0.016}$ & - & -\\
$\sigma$ (keV) & - & $9.2^{-0.3}_{+0.1}) \times 10^{-2}$ & - & -\\
$N$ (Norm. counts s$^{-1}$) & - & $(4^{-1}_{+4}) \times 10^4$ & - & -\\
$EW^\ddagger$ (eV) & - & 737 & - & -\\
\hline
{\bf O VII (2,5 $\rightarrow$ 1)} & & & & \\
$E$ (keV) & - & - & $0.568^{-0.004}_{+0.004}$ & -\\
$\sigma$ (keV) & - & - & $(6.1^{-0.3}_{+0.1}) \times 10^{-2}$ & -\\
$N$ (Norm. counts s$^{-1}$) & - & - & $(4.7^{-0.3}_{+0.7}) \times 10^{-2}$ & -\\
$EW^\ddagger$ (eV) & - & - & 198 & -\\
\hline
{\bf N VII (6,7 $\rightarrow$ 1)/O VII (2,5,6 $\rightarrow$ 1)} & & & & \\
$E$ (keV) & - & $0.597^{-0.002}_{+0.003}$ & - & -\\
$\sigma$ (keV) & - & $<0.06$ & - & -\\
$N$ (Norm. counts s$^{-1}$) & - & $(2.4^{-0.4}_{+1.5}) \times 10^5$ & - & -\\
$EW^\ddagger$ (eV) & - & 472 & - & -\\
\hline
{\bf O VIII (6,7 $\rightarrow$ 1)/Fe XVIII (4,5 $\rightarrow$ 1)} & & & & \\
$E$ (keV) & - & - & $0.769^{-0.001}_{+0.001}$ & -\\
$\sigma$ (keV) & - & - & $0.112^{-0.003}_{+0.002}$ & -\\
$N$ (Norm. counts s$^{-1}$) & - & - & $(1.78^{-0.06}_{+0.11}) \times 10^{-2}$ & -\\
$EW^\ddagger$ (eV) & - & - & 338 & -\\
\hline
{\bf Ne IX (2,5 $\rightarrow$ 1)} & & & & \\
$E$ (keV) & - & $0.91^{-0.01}_{+0.01}$ & - & -\\
$\sigma$ (keV) & - & $<0.07$ & - & -\\
$N$ (Norm. counts s$^{-1}$) & - & $7.2^{-0.6}_{+0.2}$ & - & -\\
$EW^\ddagger$ (eV) & - & 147 & - & -\\
\hline
{\bf Ne X (3,4 $\rightarrow$ 1)} & & & & \\
$E$ (keV) & - & $1.014^{-0.003}_{+0.002}$ & $1.028^{-0.001}_{+0.004}$ & -\\
$\sigma$ (keV) & - & $<0.07$ & $<0.07$ & -\\
$N$ (Norm. counts s$^{-1}$) & - & $0.37^{-0.04}_{+0.03}$ & $(5.9^{-0.3}_{+0.3}) \times 10^{-4}$ & -\\
$EW^\ddagger$ (eV) & - & 68 & 33 & -\\
\hline
{\bf Mg XI (2 $\rightarrow$ 1)} & & & & \\
$E$ (keV) & $1.346^{-0.002}_{+0.001}$ & $1.347^{-0.004}_{+0.003}$ & $1.332^{-0.002}_{+0.006}$ & $1.33^{-0.02}_{+0.02}$\\
$\sigma$ (keV) & $<0.08$ & $<0.08$ & $<0.08$ & $<0.08$\\
$N$ (Norm. counts s$^{-1}$) & $2.6^{-0.1}_{+0.1} \times 10^{-3}$ & $(2.0^{-0.1}_{+0.3}) \times 10^{-3}$ & $(2.03^{-0.08}_{+0.08}) \times 10^{-4}$ & $(1.8^{-0.3}_{+0.3}) \times 10^{-4}$\\
$EW^\ddagger$ (eV) & 95 & 337 & 62 & 84\\
\hline
\hline
\multicolumn{5}{l}{
\begin{minipage}{14cm}
$^\dagger$ The absorption column density of N49 is fitted using the LMC abundances: He=0.89, C=0.30, N=0.12, O=0.26, Ne=0.33, Na=0.30, Mg=0.32, Al=0.30, Si=0.30, S=0.31, Cl=0.31, Ar=0.54, Ca=0.34, Cr=0.61, Fe=0.36, Co=0.30 \& Ni=0.62. We have added also the galactic absorption $N_H=6 \times 10^{20} cm^{-2}$.
\end{minipage}}\\
\multicolumn{5}{l}{
\begin{minipage}{14cm}
$^\ddagger$ Equivalent Width.
\end{minipage}}
\end{tabular}
\end{center}
\caption{Summary of the fitted models for Kes 73, CTB 109, N 49 and Kes 75.}
\label{tab1_snr}
\end{table*}

\begin{table*}
\scriptsize
\begin{center}
\begin{tabular}{lllll}
\hline
\bf{Parameter} & {\bf Kes 73} & {\bf CTB 109} & {\bf N 49}$^{\dagger}$ & {\bf Kes 75}\\
\hline
{\bf Mg XII (3,4 $\rightarrow$ 1)} & & & & \\
$E$ (keV) & - & - & $1.459^{-0.005}_{+0.006}$ & -\\
$\sigma$ (keV) & - & - & $<0.08$ & -\\
$N$ (Norm. counts s$^{-1}$) & - & - & $(3.9^{-0.5}_{+0.6}) \times 10^{-5}$ & -\\
$EW^\ddagger$ (eV) & - & - & 20 & -\\
\hline
{\bf Si XIII (2,5,6,7 $\rightarrow$ 1)} & & & & \\
$E$ (keV) & $1.8521^{-0.0001}_{+0.0001}$ & $1.856^{-0.001}_{+0.006}$ & $1.848^{-0.003}_{+0.002}$ & $1.851^{-0.003}_{+0.012}$\\
$\sigma$ (keV) & $<0.02$ & $<0.02$ & $(2.3^{-0.6}_{+0.6}) \times 10^{-2}$ & $<0.02$\\
$N$ (Norm. counts s$^{-1}$) & $2.76^{-0.06}_{+0.06} \times 10^{-3}$ & $(7.0^{-0.2}_{+0.3}) \times 10^{-4}$ & $(1.68^{-0.04}_{+0.06}) \times 10^{-4}$ & $(2.6^{-0.1}_{+0.2}) \times 10^{-4}$\\
$EW^\ddagger$ (eV) & 368 & 278 & 299 & 232\\
\hline
{\bf Si XIV (3,4 $\rightarrow$ 1)} & & & & \\
$E$ (keV) & - & - & $1.998^{-0.002}_{+0.007}$ & -\\
$\sigma$ (keV) & - & - & $<0.09$ & -\\
$N$ (Norm. counts s$^{-1}$) & - & - & $(5.2^{-0.4}_{+0.3}) \times 10^{-5}$ & -\\
$EW^\ddagger$ (eV) & - & - & 132 & -\\
\hline
{\bf Si XIII (13 $\rightarrow$ 1)} & & & & \\
$E$ (keV) & $2.201^{-0.010}_{+0.009}$ & - & - & $2.21^{-0.02}_{+0.04}$\\
$\sigma$ (keV) & $<0.09$ & - & - & $<0.09$\\
$N$ (Norm. counts s$^{-1}$) & $(1.6^{-0.2}_{+0.2}) \times 10^{-4}$ & - & - & $(3.4^{-0.9}_{+1.1}) \times 10^{-5}$\\
$EW^\ddagger$ (eV) & 46 & - & - & 45\\
\hline
{\bf S XV (2,5,6,7 $\rightarrow$ 1)} & & & & \\
$E$ (keV) & $2.452^{-0.002}_{+0.002}$ & - & $2.444^{-0.005}_{+0.005}$ & $2.437^{-0.005}_{+0.007}$\\
$\sigma$ (keV) & $<0.09$ & - & $<0.09$ & $<0.09$\\
$N$ (Norm. counts s$^{-1}$) & $(8.0^{-0.3}_{+0.2}) \times 10^{-4}$ & - & $(6.8^{-0.4}_{+0.4}) \times 10^{-5}$ & $(1.09^{-0.12}_{+0.08}) \times 10^{-4}$\\
$EW^\ddagger$ (eV) & 375 & - & 299 & 178\\
\hline
{\bf S XV (13 $\rightarrow$ 1)} & & & & \\
$E$ (keV) & - & - & - & -\\
$\sigma$ (keV) & - & - & - & -\\
$N$ (Norm. counts s$^{-1}$) & - & - & - & -\\
$EW^\ddagger$ (eV) & - & - & - & -\\
\hline
{\bf Ar XVII (2,5,6,7 $\rightarrow$ 1)} & & & & \\
$E$ (keV) & $3.13^{-0.01}_{+0.01}$ & - & $3.12^{-0.02}_{+0.02}$ & -\\
$\sigma$ (keV) & $<0.1$ & - & $<0.1$ & -\\
$N$ (Norm. counts s$^{-1}$) & $(9^{-1}_{+1}) \times 10^{-5}$ & - & $(7^{-1}_{+1}) \times 10^{-6}$ & -\\
$EW^\ddagger$ (eV) & 120 & - & 110 & -\\
\hline
{\bf Fe XXV (7 $\rightarrow$ 1)} & & & & \\
$E$ (keV) & $6.7^{-0.2}_{+0.2}$ & - & - & -\\
$\sigma$ (keV) & $0.5^{-0.1}_{+0.2}$ & - & - & -\\
$N$ (Norm. counts s$^{-1}$) & $2.9^{-0.6}_{+0.6} \times 10^{-5}$ & - & - & -\\
$EW^\ddagger$ (eV) & 1890 & - & - & -\\
\hline
$\chi^2_r$ & 1.57 (985) & 2.05 (477) & 1.84 (578) & 1.12 (258)\\
\hline
\hline
\multicolumn{5}{l}{
\begin{minipage}{14cm}
$^\dagger$ The absorption column density of N49 is fitted using the LMC abundances: He=0.89, C=0.30, N=0.12, O=0.26, Ne=0.33, Na=0.30, Mg=0.32, Al=0.30, Si=0.30, S=0.31, Cl=0.31, Ar=0.54, Ca=0.34, Cr=0.61, Fe=0.36, Co=0.30 \& Ni=0.62. We have added also the galactic absorption $N_H=6 \times 10^{20} cm^{-2}$.
\end{minipage}}\\
\multicolumn{5}{l}{
\begin{minipage}{14cm}
$^\ddagger$ Equivalent Width.
\end{minipage}}
\end{tabular}
\end{center}
\caption{Continued.}
\label{tab2_snr}
\end{table*}

\begin{table*}
\scriptsize
\centering
\begin{tabular}{lllll}
\hline
 \multicolumn{5}{c}{{\tt VNEI}}\\
\hline
\bf{Parameter} & \bf{Kes 73} & \bf{CTB 109} & \bf{N 49$^\dagger$} & \bf{Kes 75}\\
\hline
$N_H$ (cm$^{-2}$) & $2.51^{-0.08}_{+0.06}$ & $0.695^{-0.018}_{+0.005}$ & $1.03^{-0.02}_{+0.02}$ & $3.71^{-0.06}_{+0.07}$\\
kT$_{brems}$ (keV) & $0.41^{-0.03}_{+0.05}$ & - & $0.99^{-0.01}_{+0.02}$ & $0.31^{-0.04}_{+0.05}$\\
N$_{brems}$ (Norm. counts s$^{-1}$) & $0.5^{-0.2}_{+0.2}$ & - & $(5.4^{-0.3}_{+0.3}) \times 10^{-3}$ & $0.4^{-0.2}_{+0.5}$\\
$kT$ (keV) & $1.51^{-0.08}_{+0.15}$ & $0.297^{-0.004}_{+0.007}$ & $0.1650^{-0.0003}_{+0.0011}$ & $2.0^{-0.1}_{+0.2}$\\
$O$ & 1 (fixed) & $0.16^{-0.02}_{+0.01}$ & $0.137^{-0.003}_{+0.002}$ & 1 (fixed)\\
$Ne$ & 1 (fixed) & $0.27^{-0.01}_{+0.01}$ & $0.175^{-0.004}_{+0.004}$ & 1 (fixed)\\
$Mg$ & $1.30^{-0.11}_{+0.09}$ & $0.23^{-0.02}_{+0.01}$ & $0.36^{-0.01}_{+0.01}$ & $0.51^{-0.08}_{+0.09}$\\
$Si$ & $1.6^{-0.1}_{0.2}$ & $0.49^{-0.05}_{+0.03}$ & 1 (fixed) & $0.56^{-0.04}_{+0.05}$\\
$S$ & $2.1^{-0.2}_{+0.4}$ & 1 (fixed) & 1 (fixed) & $0.9^{-0.1}_{+0.2}$\\
$Ar$ & $3.1^{-0.6}_{+0.9}$ & 1 (fixed) & 1 (fixed) & $1.2^{-0.6}_{+0.8}$\\
$Ca$ & $6^{-2}_{+4}$ & 1 (fixed) & 1 (fixed) & 1 (fixed)\\
$Fe$ & 1 (fixed) & $0.226^{-0.024}_{+0.008}$ & 1 (fixed) & 1 (fixed)\\
$E_1$ (keV) & - & - & $0.729^{-0.002}_{+0.005}$ & -\\
$\sigma_1$ (keV) & - & - & $<0.07$ & -\\
$N_1$ (Norm. counts s$^{-1}$) & - & - & $(5.4^{-0.3}_{+0.3}) \times 10^{-3}$ & -\\
$E_2$ (keV) & - & - & $1.018^{-0.001}_{+0.001}$ & -\\
$\sigma_2$ (keV) & - & - & $<0.07$ & -\\
$N_2$ (Norm. counts s$^{-1}$) & - & - & $(1.20^{0.04}_{0.04}) \times 10^{-3}$ & -\\
$E_3$ (keV) & - & - & $1.467^{-0.008}_{+0.004}$ & -\\
$\sigma_3$ (keV) & - & - & $<0.08$ & -\\
$N_3$ (Norm. counts s$^{-1}$) & - & - & $(4.9^{0.6}_{0.6}) \times 10^{-5}$ & -\\
$E_4$ (keV) & - & - & $1.846^{-0.003}_{+0.003}$ & -\\
$\sigma_4$ (keV) & - & - & $<0.09$ & -\\
$N_4$ (Norm. counts s$^{-1}$) & - & - & $(1.56^{0.07}_{0.07}) \times 10^{-4}$ & -\\
$E_5$ (keV) & - & - & $1.998^{-0.003}_{+0.028}$ & -\\
$\sigma_5$ (keV) & - & - & $<0.09$ & -\\
$N_5$ (Norm. counts s$^{-1}$) & - & - & $(5.3^{0.5}_{0.5}) \times 10^{-5}$ & -\\
$E_6$ (keV) & - & - & $2.445^{-0.005}_{+0.005}$ & -\\
$\sigma_6$ (keV) & - & - & $<0.1$ & -\\
$N_6$ (Norm. counts s$^{-1}$) & - & - & $(6.4^{-0.4}_{+0.3}) \times 10^{-5}$ & -\\
$E_7$ (keV) & - & - & $3.12^{-0.02}_{+0.02}$ & -\\
$\sigma_7$ (keV) & - & - & $<0.1$ & -\\
$N_7$ (Norm. counts s$^{-1}$) & - & - & $(7^{-1}_{+1}) \times 10^{-6}$ & -\\
$\tau$ (s cm$^{-3}$) & $(5.1^{-0.8}_{+0.6}) \times 10^{10}$ & $(6.7^{-1.0}_{+0.8}) \times 10^{11}$ & $(1.3^{-0.2}_{+0.1}) \times 10^{12}$ & $(2.4^{-0.3}_{+0.3}) \times 10^{10}$\\
$N$ (Norm. counts s$^{-1}$) & $(3.9^{-0.9}_{+0.6}) \times 10^{-2}$ & $0.35^{-0.04}_{+0.02}$ & $1.69^{-0.02}_{+0.03}$ & $0.021^{-0.003}_{+0.003}$\\
$\chi^2_r$ & 1.56 (997) & 2.60 (491) & 1.87 (569) & 1.19 (236)\\
\hline
\hline
\end{tabular}
\caption{Fits for Kes 73, CTB 109, N 49 \& Kes 75 using a {\tt vnei} plasma model. A second thermal Bremsstrahlung component
is included in some cases.$^\dagger$ The absorption column density of N49 is fitted using the LMC abundances: He=0.89, C=0.30,
N=0.12, O=0.26, Ne=0.33, Na=0.30, Mg=0.32, Al=0.30, Si=0.30, S=0.31, Cl=0.31, Ar=0.54, Ca=0.34, Cr=0.61, Fe=0.36, Co=0.30 \& Ni=0.62.
We have added also the galactic absorption $N_H=6 \times 10^{20} cm^{-2}$.}
\label{plasmamodels}
\end{table*}

\begin{table*}
\scriptsize
\centering
\begin{tabular}{l@{\ \ }l@{\ }l@{\ }l@{\ \ }l@{\ }l@{\ \ }l@{\ \ }l@{\ \ }l}
\hline
 \multicolumn{9}{c}{SNRs with magnetars}\\
\hline
\bf{Name} & \bf{Central source} & \bf{Distance} &  \bf{Radius} & \bf{Age} & \bf{$\dot{E}$} & \bf{B$_s$} & \bf{F$_X$} & \bf{L$_X$}\\
 & & \bf{(kpc)} & \bf{(pc)} & \bf{(kyr)} & \bf{(erg s$^{-1}$)} & \bf{(G)} & \bf{(erg cm$^{-2}$ s$^{-1}$)} & \bf{(erg s$^{-1}$)}\\
\hline
Kes 75 & J1846-0258 [26] & 10.6 [66] & 5.5$^{-0.3}_{+0.3}$ [10] & 0.9 [6] & $8.06 \times 10^{36}$ [40] & $4.88 \times 10^{13}$ [40] & $2.69 \times 10^{-10}$ & $3.61 \times 10^{36}$\\
Kes 73 & 1E 1841-045 [72] & 6.7$^{-1.0}_{+1.8}$ [61] &  4.5$^{-0.1}_{+0.1}$ [10] & 1.3$^{-0.2}_{+0.2}$ [73] & $1.08 \times 10^{33}$ [31] & $7.34 \times 10^{14}$ [31] & $4.39 \times 10^{-10}$ & $2.36^{-0.65}_{+1.43} \times 10^{36}$\\
N 49 & RX J0526-6604 [36] & 50 [36] & 20.4 [10] & 4.8 [48] & $4.92 \times 10^{33}$ [36] & $7.32 \times 10^{14}$ [36] & $2.41 \times 10^{-10}$ & $7.21 \times 10^{37}$\\
CTB 109 & 1E 2259+586 [2] & 3$^{-0.5}_{+0.5}$ [33] & 12.6$^{-1.3}_{+1.3}$ [10] & 14$^{-2}_{+2}$ [62] & $5.54 \times 10^{31}$ [2] & $5.84 \times 10^{13}$ [2] & $1.94 \times 10^{-10}$ & $2.09^{-0.64}_{+0.75} \times 10^{35}$\\
\hline
\hline
 \multicolumn{9}{c}{SNRs with CCOs}\\
\hline
Cas A & CXO J2323+5848 [45] & 3.4$^{-0.1}_{+0.3}$ [53] & 2.8$^{-0.1}_{+0.1}$ [10] & 0.326$^{-27}_{+27}$ [17] & - & - & $2.06 \times 10^{-8}$ [10] & $2.85^{-0.20}_{+0.50} \times 10^{37}$ [10]\\
G350.1-0.3 & XMMU J1720-3726 [23] & 4.5 [23] & 2.5$^{-0.4}_{+0.4}$ [10] & 0.9 [23] & - & - & $1.64 \times 10^{-9}$ [10] & $3.97 \times 10^{36}$ [10]\\
G330.2+1.0 & CXOU J1601-5133 [47] & 4.9$^{-0.3}_{+0.3}$ [53] & 7.8$^{-0.8}_{+0.8}$ [10] & 1.1 [47] & - & - & $1.60 \times 10^{-11}$ [71] & $4.60^{-0.55}_{+0.57} \times 10^{34}$ [71]\\
G347.3-0.5 & 1 WGA J1713-3949 [38] & 1 [34] & 8.7$^{-0.8}_{+0.8}$ [18] & 1.6 [18] & - & - & $4.40 \times 10^{-10}$ [51] & $5.26 \times 10^{34}$\\
Vela Jr. & CXOU J0852-4617 [49] & 0.75$^{-0.55}_{+0.25}$ [31] &  13.1 [10] & 1.7$^{-0}_{+2.6}$ [31] & - & - & $8.30 \times 10^{-11}$ [1] & $5.58^{-3.10}_{+4.34} \times 10^{33}$\\
RCW 103 & 1E 1613-5055 [25] & 3.1 [55] &  4.1$^{-0.1}_{+0.1}$ [10] & 2 [7] & - & - & $1.70 \times 10^{-8}$ [10] & $1.95 \times 10^{37}$\\
G349.7+0.2 & CXOU J1718-3726 [39] & 22.4 [20] &  8.2 [39] & 3.5 [39] & - & - & $6.50 \times 10^{-10}$ [39] & $3.90 \times 10^{37}$\\
Puppis A & RX J0822-4300 [4] & 2.2$^{-0.3}_{+0.3}$ [54] &  17.5$^{-1.7}_{+1.7}$ [16] & 4.45$^{-0.75}_{+0.75}$ [4] & - & - & $2.16 \times 10^{-8}$ & $1.20^{0.90}_{+1.55} \times 10^{37}$ [14]\\
Kes 79 & J1852+0040 [63] & 7.1 [8] &  9.2$^{-1.0}_{+1.0}$ [10] & 6.0$^{-0.2}_{+0.4}$ [67] & $2.96 \times 10^{32}$ [27] & $3.05 \times 10^{10}$ [27] & $4.64 \times 10^{-10}$ [67] & $2.80 \times 10^{36}$ [67]\\
G296.5+10.0 & 1E 1207-5209 [24] & 2.1$^{-0.8}_{+1.8}$ [24] &  24.8 [32] & 7 [57] & $9.58 \times 10^{33}$ [50] & $2.83 \times 10^{12}$ [50] & $1.67 \times 10^{-9}$ [44] & $8.81^{-5.40}_{+21.60} \times 10^{34}$\\
\hline
\hline
 \multicolumn{9}{c}{SNRs with high-B PSRs}\\
\hline
MSH 15-52 & J1513-5908 [21] & 5.2$^{-1.4}_{+1.4}$ [15] & 22.7 [46] & 1.9 [15] & $1.75 \times 10^{37}$ [41] & $1.54 \times 10^{13}$ [41] & $7.80 \times 10^{-11}$ [46] & $2.52^{-1.17}_{+1.54} \times 10^{35}$\\
MSH 11-54 & J1124-5916 [29] & 6.2$^{-0.9}_{+0.9}$ [22] & 16.2$^{-0.2}_{+0.2}$ [10] & 2.99$^{-0.06}_{+0.06}$ [76] & $1.19 \times 10^{37}$ [52] & $1.02 \times 10^{13}$ [52] & $2.09 \times 10^{-9}$ [10] & $9.61^{-2.59}_{+2.99} \times 10^{36}$\\
G292.2-0.5 & J1119-6127 [37] & 8.4$^{-0.4}_{+0.4}$ [9] &  21.1$^{-3.8}_{+3.8}$ [10] & 7.1$^{-0.2}_{+0.5}$ [37] & $2.34 \times 10^{36}$ [75] & $4.10 \times 10^{13}$ [75] & $1.98 \times 10^{-11}$ [37] & $1.67^{-0.15}_{+0.16} \times 10^{35}$\\
\hline
\hline
 \multicolumn{9}{c}{SNRs with normal PSRs}\\
\hline
G21.5-0.9 & J1833-1034 [43] & 4.7$^{-0.4}_{+0.4}$ [69] & 3.2$^{-0.1}_{+0.1}$ [10] & 0.87$^{-1.5}_{+2.0}$ [5] & $3.37 \times 10^{37}$ [58] & $3.58 \times 10^{12}$ [58] & $6.69 \times 10^{-13}$ & $1.77^{-0.31}_{0.29} \times 10^{33}$ [43]\\
G11.2-0.3 & J1811-1925 [70] & 5 [30] & 2.3$^{-0.1}_{+0.1}$ [10] & 1.616 [68] & $6.42 \times 10^{36}$ [70] & $1.71 \times 10^{12}$ [70] & $3.98 \times 10^{-9}$ [10] & $1.19 \times 10^{37}$ [10]\\
G8.7-0.1 & J1803-2137 [19] & 4 [19] & 29.1 [19] & 15$^{-6}_{+6}$ [19] & $2.22 \times 10^{36}$ [77] & $4.92 \times 10^{12}$ [77] & $2.00 \times 10^{-10}$ [19] & $3.83 \times 10^{35}$\\
Vela & J0835-4510 [3] & 0.287$^{-0.017}_{+0.019}$ [13] & 20.1 [42] & 18$^{-9}_{+9}$ [3] & $6.92 \times 10^{36}$ [12] & $3.38 \times 10^{12}$ [12] & $2.94 \times 10^{-8}$ & $2.90^{-0.34}_{+0.39} \times 10^{35}$ [42]\\
MSH 11-61A & J1105-6107 [64] & 7 [64] & 12.1$^{-2.2}_{+2.2}$ [64] & 20$^{-5}_{+5}$ [64] & $2.48 \times 10^{36}$ [74] & $1.01 \times 10^{12}$ [74] & $8.06 \times 10^{-11}$ [10] & $4.71 \times 10^{35}$ [10]\\
W 44 & J1856+0113 [11] & 2.5 [11] & 10.8$^{-2.0}_{+2.0}$ [11] & 20$^{-4}_{+4}$ [11] & $4.30 \times 10^{35}$ [28] & $7.55 \times 10^{12}$ [28] & $1.80 \times 10^{-9}$ [56] & $1.35 \times 10^{36}$\\
CTB 80 & J1952+3252 [60] & 2 [65] & 1.5 [60] & 51 [78] & $3.74 \times 10^{36}$ [28] & $4.86 \times 10^{11}$ [28] & $2.40 \times 10^{-12}$ & $1.15 \times 10^{33}$ [59]\\
\hline
\hline
\end{tabular}
\caption{SNRs considered in our X-ray luminosity analysis. The data without references is extracted from this work or deduced from the data obtained in the literature. The
references are: $^{[1]}$\protect\citet{aha07}, $^{[2]}$\protect\citet{archi13}, $^{[3]}$\protect\citet{asch95}, $^{[4]}$\protect\citet{becker12}, $^{[5]}$\protect\citet{bieten08},
$^{[6]}$\protect\citet{blan96}, $^{[7]}$\protect\citet{carter97}, $^{[8]}$\protect\citet{case98}, $^{[9]}$\protect\citet{caswell04},
$^{[10]}${\it Chandra} SNR catalog\protect\footnote{\tt{http://hea-www.cfa.harvard.edu/ChandraSNR/}}, $^{[11]}$\protect\citet{cox99}, $^{[12]}$\protect\citet{dodson02},
$^{[13]}$\protect\citet{dodson03}, $^{[14]}$\protect\citet{dubner13}, $^{[15]}$\protect\citet{fang10}, $^{[16]}$\protect\citet{ferrand12}, $^{[17]}$\protect\citet{fesen06},
$^{[18]}$\protect\citet{fesen12}, $^{[19]}$\protect\citet{fin94}, $^{[20]}$\protect\citet{frail96}, $^{[21]}$\protect\citet{gaen99}, $^{[22]}$\protect\citet{gaen03},
$^{[23]}$\protect\citet{gaen08}, $^{[24]}$\protect\citet{giac00}, $^{[25]}$\protect\citet{gott99a}, $^{[26]}$\protect\citet{gott00}, $^{[27]}$\protect\citet{halpern10},
$^{[28]}$\protect\citet{hobbs04}, $^{[29]}$\protect\citet{hughes03}, $^{[30]}$\protect\citet{kaspi01}, $^{[31]}$\protect\citet{kat08}, $^{[32]}$\protect\citet{kellett87},
$^{[33]}$\protect\citet{kothes02}, $^{[34]}$\protect\citet{koyama97}, $^{[31]}$\protect\citet{kuiper06}, $^{[36]}$\protect\citet{kulk03}, $^{[37]}$\protect\citet{kumar12},
$^{[38]}$\protect\citet{lazen03}, $^{[39]}$\protect\citet{lazen05}, $^{[40]}$\citet{liv11a}, $^{[41]}$\protect\citet{liv11b}, $^{[42]}$\protect\citet{lu00}, $^{[43]}$\protect\citet{mat10},
$^{[44]}$\protect\citet{mat88}, $^{[45]}$\protect\citet{mere02}, $^{[46]}$\protect\citet{mineo01}, $^{[47]}$\protect\citet{park09}, $^{[48]}$\protect\citet{park12},
$^{[49]}$\protect\citet{pav01}, $^{[50]}$\protect\citet{pav02}, $^{[51]}$\protect\citet{pfe96}, $^{[52]}$\protect\citet{ray11}, $^{[53]}$\protect\citet{reed95},
$^{[54]}$\protect\citet{rey95}, $^{[55]}$\protect\citet{rey04}, $^{[56]}$\protect\citet{rho94}, $^{[57]}$\protect\citet{roger88}, $^{[58]}$\protect\citet{roy12},
$^{[59]}$\protect\citet{safi94}, $^{[60]}$\protect\citet{safi95}, $^{[61]}$\protect\citet{san92}, $^{[62]}$\protect\citet{sasaki13}, $^{[63]}$\protect\citet{seward03},
$^{[64]}$\protect\citet{slane02}, $^{[65]}$\protect\citet{strom00}, $^{[66]}$\protect\citet{su09}, $^{[67]}$\protect\citet{sun04}, $^{[68]}$\protect\citet{tam03},
$^{[69]}$\protect\citet{tian08a}, $^{[70]}$\protect\citet{torii99}, $^{[71]}$\protect\citet{torii06}, $^{[72]}$\protect\citet{vasisht97}, $^{[73]}$\protect\citet{vink06},
$^{[74]}$\protect\citet{wang00}, $^{[75]}$\protect\citet{welt11}, $^{[76]}$\protect\citet{wink09}, $^{[77]}$\protect\citet{yuan10}, $^{[78]}$\protect\citet{zeiger08}.}
\label{snrdata}
\end{table*}


\begin{table*}
\scriptsize
\centering
\begin{tabular}{@{}lcccccccc}
\hline
{\bf SNR} & {\bf Galaxy} & {\bf Age (yr)} & \multicolumn{6}{c}{{\bf Element}} \\
 & & & {\it O VII} & {\it O VIII} & {\it O VIII} & {\it Ne IX} & {\it Ne X} & {\it Ne X}\\
 & & & {\it (2,5,7 $\to$ 1)} & {\it (3,4 $\to$ 1)} & {\it (6,7 $\to$ 1)} & {\it (2,5 $\to$ 1)} & {\it (3,4 $\to$ 1)} & {\it (6,7 $\to$ 1)}\\
 & & & {\it (0.574 KeV)} & {\it (0.653 KeV)} & {\it (0.774 KeV)} & {\it (0.915 KeV)} & {\it (1.022 KeV)}& {\it (1.21 KeV)} \\
\hline
Kes73 & MW & 1100-1500 & & & & & &\\
CTB109 & MW & 7900-9700 & & & & X & X &\\
Kes75 & MW & 900-4300 & & & & & &\\
N49 & LMC & 5000 & X & & X & X & X &\\
\hline
G1.9+1.3 [2] & MW & 110-170 & & & & & &\\
Kepler [3],[8],[12] & MW & 408 & & & & X & &\\
Tycho [4],[5],[6],[13] & MW & 440 & & & & X & &\\
SN1006 [10],[19] & MW & 1006 & X & & X & & & X\\
Cas A [1],[9],[16] & MW & 316-352 & X & X & X & X & X &\\
MSH11-54 [11],[14] & MW & 2930-3050 & X & X & X & X & X &\\
Puppis A [7],[17],[18] & MW & 3700-5500 & X & X & X & X & X & X\\
B0509-67.5 [15] & LMC & 400 & X & X & & X & &\\
\hline
\hline
 & & & {\it Mg XI} & {\it Mg XII} & {\it Si XIII} & {\it Si XIV} & {\it Si XIII} & {\it S XV}\\
 & & & {\it (2,5,6,7 $\to$ 1)} & {\it (3,4 $\to$ 1)} & {\it (2,5,6,7 $\to$ 1)} & {\it (3,4 $\to$ 1)} & {\it (13 $\to$ 1)} & {\it (2,5,6,7 $\to$ 1)}\\
 & & & {\it (1.35 KeV)} & {\it (1.47 KeV)} & {\it (1.86 KeV)} & {\it (2.00 KeV)} & {\it (2.18 KeV)} & {\it (2.46 KeV)}\\
\hline
Kes73 & MW & 1100-1500 & X & & X & & X & X\\
CTB109 & MW & 7900-9700 & X & & X & & &\\
Kes75 & MW & 900-4300 & X & & X & & X & X\\
N49 & LMC & 5000 & X & X & X & X & & X\\
\hline
G1.9+1.3 & MW & 110-170 & X & & X & & & X\\
Kepler & MW & 408 & X & & X & X & X & X\\
Tycho & MW & 440 & & & X & X & X & X\\
SN1006 & MW & 1006 & X & & X & & &\\
Cas A & MW & 316-352 & X & X & X & X & X & X\\
MSH11-54 & MW & 2930-3050 & X & X & X & & & X\\
Puppis A & MW & 3700-5500 & X & & X & & X & X\\
B0509-67.5 & LMC & 400 & X & & X & & X & X\\
\hline
\hline
 & & & {\it S XV} & {\it Ar XVII} & {\it Ca XIX} & {\it Fe XXV}\\
 & & & {\it (13 $\to$ 1)} & {\it (2,5,6,7 $\to$ 1)} & {\it (2,5,6,7 $\to$ 1)} & {\it K-shell}\\
 & & & {\it (2.88 KeV)} & {\it (3.13 KeV)} & {\it (3.89 KeV)} & {\it (6.65 KeV)}\\
\hline
Kes73 & MW & 1100-1500 & & X & & X\\
CTB109 & MW & 7900-9700 & & & & \\
Kes75 & MW & 900-4300 & & & & \\
N49 & LMC & 5000 & & X & & \\
\hline
G1.9+1.3 & MW & 110-170 & & X & X & X\\
Kepler & MW & 408 & X & X & X & X\\
Tycho & MW & 440 & X & X & X & X\\
SN1006 & MW & 1006 & & & & \\
Cas A & MW & 316-352 & X & X & X & X\\
MSH11-54 & MW & 2930-3050 & & & & \\
Puppis A & MW & 3700-5500 & & & & \\
B0509-67.5 & LMC & 400 & & X & X & X\\
\hline
\hline
\end{tabular}
\caption{Summary of line detections in X-ray for some important SNRs compared with lines detected in our analysis. The references are:
$^{[1]}$\protect\citet{blee01}, $^{[2]}$\protect\citet{borkowski10}, $^{[3]}$\protect\citet{cassam04}, $^{[4]}$\protect\citet{decour01},
$^{[5]}$\protect\citet{hayato10}, $^{[6]}$\protect\citet{hwang97}, $^{[7]}$\protect\citet{hwang08}, $^{[8]}$\protect\citet{kinugasa99},
$^{[9]}$\protect\citet{maeda09}, $^{[10]}$\protect\citet{miceli09}, $^{[11]}$\protect\citet{park07}, $^{[12]}$\protect\citet{rey07},
$^{[13]}$\protect\citet{tama09}, $^{[14]}$\protect\citet{vink04}, $^{[15]}$\protect\citet{warren04}, $^{[16]}$\protect\citet{willi02},
$^{[17]}$\protect\citet{winkler81a}, $^{[18]}$\protect\citet{winkler81b}, $^{[19]}$\protect\citet{yama08}}
\label{tab:snr_list}
\end{table*}


\end{document}